\documentclass[a4paper,scriptaddress,preprint,prd,floatfix]{revtex4-1}
\usepackage{amsmath}
\usepackage{palatino,url,amssymb,graphicx,epic,amsmath,xy,epsfig}
\usepackage[toc,page]{appendix}
\usepackage[margin=2cm]{geometry}
\usepackage{comment}
\usepackage{float}

\usepackage[T1]{fontenc}
\begin{document}

\title{\textit{MasQU}: Finite Differences on Masked Irregular Stokes Q, U Grids}

\author{Jude Bowyer$^{*}$}
\author{Andrew H. Jaffe$^{\dag}$}
\author{Dmitri I. Novikov$^{\ddag}$}
\affiliation{Astrophysics Group, Blackett Laboratory, Imperal College, \\ Prince Consort Road, London, SW7 2BZ, U.K.}

\begin{abstract}
\noindent The detection of $B$-mode polarization in the CMB is one of the most important outstanding tests of inflationary cosmology. One of the necessary steps for extracting polarization information in the CMB is reducing contamination from so-called ``ambiguous modes'' on a masked sky, which contain leakage from the larger $E$-mode signal. This can be achieved by utilising derivative operators on the real-space Stokes $Q$ and $U$ parameters. This paper presents an algorithm and a software package to perform this procedure on the nearly full sky, i.e., with projects such as the \textit{Planck Surveyor} and future satellites in mind; in particular, the package can perform finite differences on masked, irregular grids and is applied to a semi-regular spherical pixellization, the \textit{HEALPix} grid. The formalism reduces to the known finite-difference solutions in the case of a regular grid. We quantify full-sky improvements on the possible bounds on the CMB $B$-mode signal. We find that in the specific case of $E$ and $B$-mode separation, there exists a ``pole problem'' in our formalism which produces signal contamination at very low multipoles $l$. Several solutions to the ``pole problem'' are presented; one proposed solution facilitates a calculation of a general Gaussian quadrature scheme, which finds application in calculating accurate harmonic coefficients on the HEALPix sphere. Nevertheless, on a masked sphere the software represents a considerable reduction in $B$-mode noise from limited sky coverage.
\end{abstract}

\maketitle

\section{Introduction}

\noindent The detection of primordial gravitational waves in the Cosmic Microwave Background (CMB) would be the ``smoking gun'' of inflationary cosmology [1, 2]; the existence of such gravitational waves in the inflationary scenario implies the observation of primordial tensor pertubations in the microwave sky in the form of curl-type (``$B$-mode'') polarization. When using harmonic methods on a masked sky, signal-mixing between the curl- and divergence-type (``$E$-mode'') polarization occurs [3], as the $E$- and $B$-mode power contributions from the masked regions are unknown and the signal decomposition into polarization modes is inherently non-local (hence the contributions from masked regions and their boundaries are also known as ``ambiguous modes''). Since the cosmological $B$-modes are realistically expected to be at least an order of magnitude smaller than the $E$-modes [4], such signal-mixing could potentially either drown out any true $B$-mode signal or masquerade as a false positive detection. Instead, one may perform such full-sky masked extractions by utilising real-space derivative operators on local analogues of the $E$- and $B$-modes via \textit{finite differences} [5] and thus avoid sampling in the masked regions at all. The finite-differencing method is a standard method for performing numerical derivatives, where derivatives are estimated by attaching numerical weights to the input data at a set of pixels surrounding each individual pixel.
\\The most popular pixellization for CMB analysis is the semi-regular spherical HEALPix grid [6]; standard HEALPix methods for extracting $E$- and $B$-modes are based on harmonic techniques and thus suffer from signal-mixing when a mask is present [7], since harmonic methods sample globally. Performing numerical derivatives on irregular grids, over a range of geometries, is a problem which is both general and non-trivial. While other, more regular, grids such as GLESP [8] exist for CMB analysis, they do not hold the same computational advantages as HEALPix.
\\We present the \textit{MasQU} (\textit{Mas}ked Stokes \textit{Q}, \textit{U} analysis) software which can perform such derivatives via finite-differencing. The numerical weights are not derived for particular grid types and derivative orders as in the standard finite-difference approaches, since a completely general method to calculate them algorithmically is developed. The algorithm is then tested on a Cartesian grid, and on the sphere (as supplied by the HEALPix scheme). The sphere is known for being problematic due to the poles; it will be shown that the ($\theta$, $\phi$) coordinate singularity requires special treatment for performing the finite-difference algorithm. Even after this is dealt with, in the particular case of extracting $E$ and $B$-modes there exists an accuracy problem related to the pixel positions in the HEALPix polar cap.
\\We find that, compared with standard harmonic techniques our method reduces leakage between $E$- and $B$-modes by at least an order of magnitude, the reduction increasing with multipole scale.
\\The paper proceeds as follows: in Section II the CMB polarization formalism is summarized and we develop the finite-difference weight-generation scheme. Section III then presents an implementation on the HEALPix spherical grid --- including discussions of some of the pitfalls in attempting to perform accurate derivatives at the pole. The performance of the software on the HEALPix grid in the presence of both noise and masking is analysed in Section IV, in comparison with the standard harmonic methods. We also make mock calculations for EBEX-type experiments. Finally, we conclude with some comments in Section V.

\section{Analyzing CMB Polarization on the Full Sky}

\noindent Standard inflationary scenarios in cosmology predict the existence of tensor fluctuations in the modern Universe. Tensor fluctuations from inflation are ``frozen-in'' to the CMB on superhorizon scales; this means that if primordial $B$-modes exist they will be most prominent at small values for the multipole $l$, corresponding to large angular scales. In order to fully probe large angular scales we need a full-sky analysis, something the \textit{Planck Surveyor} [9] will provide data for. In fact, whilst $E$-modes have been detected since 2002 (DASI [10]), the cosmological $B$-modes have remained elusive --- the current state of detection from Wilkinson Microwave Anisotropy Probe (WMAP) analysis giving a tensor-to-scalar ratio \begin{math}r<0.13\end{math} at 95\% confidence [11]. In the following subsection we describe our formalism for decomposing the measured CMB polarization into parameters useful for cosmology.

\subsection{CMB Polarization Formalism}

\noindent The Stokes parameters $Q$ and $U$, which describe the linear polarization of light, are dependent on the choice of reference basis but can be reparameterized in terms of two basis-independent polarization modes known as the ``E'' and ``B'' modes [12] by analogy with electromagnetism, since any tensor field of rank greater than zero can be decomposed into ``gradient''-type and ``curl''-type parts. The relation between the two sets of parameters on the spherical surface is given by
\begin{equation}\begin{split}Q(\hat{n})=-\frac{1}{2\displaystyle}\sum_{lm}\left(a^{E}_{lm}\left[\phantom{}_{2}Y_{lm}(\hat{n})+\phantom{}_{-2}Y_{lm}(\hat{n})\right]+
ia^{B}_{lm}\left[\phantom{}_{2}Y_{lm}(\hat{n})-\phantom{}_{-2}Y_{lm}(\hat{n})\right]\right)\hspace{2mm}
\\U(\hat{n})=-\frac{1}{2\displaystyle}\sum_{lm}\left(a^{B}_{lm}\left[\phantom{}_{2}Y_{lm}(\hat{n})+\phantom{}_{-2}Y_{lm}(\hat{n})\right]+
ia^{E}_{lm}\left[\phantom{}_{2}Y_{lm}(\hat{n})-\phantom{}_{-2}Y_{lm}(\hat{n})\right]\right),\end{split}\end{equation}
where the \begin{math}\phantom{}_{s}Y_{lm}\end{math} are spin-weighted spherical harmonics [13] with integer spin weight $s$ --- reducing to the standard spherical harmonics $Y_{lm}$ for $s=0$ --- $\hat{n}$ is the chosen coordinate basis and the $a^{X}_{lm}$ are the $s=0$ harmonic coefficients of $X=\{E$, $B\}$. Since one can decompose vector and tensor fields into curl and divergence parts, and cosmological vector fields decay exponentially in the inflationary scenario, the curl part of the decomposition of tensor perturbations corresponds uniquely to the $B$-modes. In the real universe of course, the situation is confused by the presence of lensing, reionization and other astrophysical phenomena restricting the detection of primordial tensor mode phenomena to low-$l$ multipoles.
\\To facilitate a real-space calculation, we can construct scalar and pseudo-scalar fields corresponding to the $E$ and $B$-modes respectively
\begin{equation}e(\hat{n})=\displaystyle\sum_{lm}\sqrt{\frac{(l-2)!}{(l+2)!}}a^{E}_{lm}Y_{lm}(\hat{n}),\indent b(\hat{n})=\displaystyle\sum_{lm}\sqrt{\frac{(l-2)!}{(l+2)!}}a^{B}_{lm}Y_{lm}(\hat{n})\end{equation}
and then relate the bi-Laplacians $\nabla^{4}\equiv\nabla^{2}(\nabla^{2}+2)$ of these fields to the Stokes parameters:
\begin{equation}\nabla^{4}e=-\frac{1}{2}[\bar{\eth}^{2}(Q+iU)+\eth^{2}(Q-iU)],\indent\nabla^{4}b=\frac{i}{2}[\bar{\eth}^{2}(Q+iU)-\eth^{2}(Q-iU)]\end{equation}
where the $\eth$ and $\bar{\eth}$ terms are respectively the Newman-Penrose spin-weight-raising and spin-weight-lowering operators, we have dropped the $\hat{n}$ for convenience, while $\eth^{n}=\overbrace{\eth\cdots\eth}^{\times n}$ and $\bar{\eth}^{n}=\overbrace{\bar{\eth}\cdots\bar{\eth}}^{\times n}$. The Newman-Penrose operators can be written
\begin{equation}\eth_{s}=-(\partial_{\theta}+i\csc\theta\partial_{\phi}-s\cot\theta),\indent\bar{\eth}_{s}=-(\partial_{\theta}-i\csc\theta\partial_{\phi}+s\cot\theta)\end{equation}
where we have appended an $s$ to the standard notation which allows easier spin-weight-counting of the entity operated on; the operation of $\eth_{s}$ on an entity with spin-weight $s$ returns a quantity with spin-weight $s+1$, and correspondingly the operation of $\bar{\eth}$ returns an entity with spin-weight $s-1$. Rearranging Eq. (3) gives
\begin{equation}\nabla^{4}e=-D^{+}_{\mp2}Q-D^{-}_{\mp2}U,\indent\nabla^{4}b=D^{-}_{\mp2}Q-D^{+}_{\mp2}U,\end{equation}
where the derivative operators are defined as
\begin{equation}\begin{split}D^{+}_{\mp2}=\frac{\eth_{-1}\eth_{-2}+\bar{\eth}_{1}\bar{\eth}_{2}}{2}=-2-\csc^{2}\theta\partial_{\phi\phi}+3\cot\theta\partial_{\theta}+\partial_{\theta\theta}
\\D^{-}_{\mp2}=\frac{\eth_{-1}\eth_{-2}-\bar{\eth}_{1}\bar{\eth}_{2}}{2i}=2\csc\theta(\cot\theta\partial_{\phi}+\partial_{\theta\phi})\hspace{0.76in}\end{split}\end{equation}
which can be used to extract $e$ and $b$ from the underlying Stokes polarization maps, with the $\mp s$ referring to the combination of spin-weights of the ($\eth$, $\bar{\eth}$) operators which define the operation (i.e., $D_{\pm s}$ would feature a combination of $\eth_s$ and $\bar{\eth}_{-s}$). The ``inverse'' relations for forming Stokes fields from the underlying $e$, $b$ fields are
\begin{equation}Q=-D^{+}_{0}e+D_{0}^{-}b,\indent U=-D^{+}_{0}b-D^{-}_{0}e,\end{equation}
so the bi-Laplacian can be written $\nabla^{4}=D^{+}_{\mp2}D^{+}_{0}+D^{-}_{\mp2}D^{-}_{0}$ and on the sphere
\begin{equation}D^{+}_{0}=\frac{\eth_{1}\eth_{0}+\bar{\eth}_{-1}\bar{\eth}_{0}}{2},\indent D^{-}_{0}=\frac{\eth_{1}\eth_{0}-\bar{\eth}_{-1}\bar{\eth}_{0}}{2i}.\end{equation}
In the flat-sky small-angle approximation, since the spin-weight $s$ of a function $\phantom{}_{s}f$ is defined by its transformation on rotation about the pole of a tangent space, via
\begin{equation}\phantom{}_{s}f\rightarrow e^{is\psi}\phantom{}_{s}f,\end{equation}
all spin-dependence of the operators vanishes
\begin{equation}\eth\rightarrow-(\partial_{x}+i\partial_{y}),\indent\bar{\eth}\rightarrow-(\partial_{x}-i\partial_{y})\end{equation}
yielding the approximation
\begin{equation}D^{+}_{\mp2}\rightarrow D^{+}_{0}\approx\partial_{xx}-\partial_{yy},\indent D^{-}_{\mp2}\rightarrow D^{-}_{0}\approx2\partial_{xy}.\end{equation}
For these operators, the relations
\begin{equation}\begin{split}[D_{0}^{+},D_{0}^{-}]=0\indent\mbox{ (flat sky)}\hspace{0.5mm}
\\D^{+}_{\mp2}D^{-}_{0}-D^{-}_{\mp2}D^{+}_{0}=0\indent\mbox{ (full sky)}\end{split}\end{equation}
are satisfied; the degree to which their discrete analogues violate these relations is a measure of signal leakage from pixellization. Since it can be shown that the harmonic-space contribution of the bi-Laplacian is
\begin{equation}\nabla^{4}\rightleftharpoons\frac{(l+2)!}{(l-2)!},\end{equation}
then relating the power spectra of the scalar and pseudo-scalar fields to that of the $E$- and $B$-modes is trivially
\begin{equation}C_{l}^{\nabla^{4}e}=\frac{(l+2)!}{(l-2)!}C_{l}^{E},\indent C_{l}^{\nabla^{4}b}=\frac{(l+2)!}{(l-2)!}C_{l}^{B}.\end{equation}

\subsection{Computing Derivatives by Finite-Differencing}

\noindent A derivative (of order $m$) of a function $f(x)$ with respect to coordinate $x$ at a given pixel labelled $i$ can be computed as the sum of weighted values of the function at a surrounding sample of pixels $j$ (the pixel `stencil') [5]:
\begin{equation}\partial_{x^{m}}f_{i}\approx\sum_{j}^{\mathrm{pixels}}w^{(m)}_{ij}f_{j}\end{equation}
where $w$ is the weight matrix. The finite-difference method involves estimating derivatives by taking weighted differences of function values at selected pixels (see Appendix A for further details); a general method [14] for computing the weights on an arbitrary 1-d grid requires, for the derivatives at a pixel $i$, solving the linear system
\begin{equation}\left(\begin{array}{ccc}\Delta_{i,1}^{0} & \cdots & \Delta_{i,n}^{0} \\ \vdots &  & \vdots \\ \Delta_{i,1}^{n-1} & \cdots & \Delta_{i,n}^{n-1}\end{array} \right)\left(\begin{array}{c}w^{(m)}_{i,1} \\ \vdots \\ w^{(m)}_{i,n}\end{array}\right)=\left(\begin{array}{c}0!\delta_{m,0} \\ \vdots \\ (n-1)!\delta_{m,n-1}\end{array}\right)\end{equation}
with $\delta_{mn}$ a Kronecker delta for separating the $m^{\mathrm{th}}$ derivative such that only a single entry on the right-hand side is nonzero for the given derivative, and $\Delta_{ij}=x_{j}-x_{i}$ a position difference term. Various tests of this method can be found in Fig. 1.
\\The geometric array in Eq. (16) is the differenced ``Vandermonde'' matrix [15]. The Vandermonde matrix is related to the Lagrange basis interpolating polynomial 
\begin{equation}L_{i}(x)=\prod_{1\leq j\leq n; j\neq i}\frac{x-x_{j}}{x_{i}-x_{j}}\end{equation}
which can be used to interpolate a function $f$ sampled at $n$ points by
\begin{equation}f_{i}\approx\sum_{j}^{n}f_{j}L_{j}(x_{i}).\end{equation}
In the Vandermonde formalism this polynomial can be identified with the determinants of arrays determined by the pixel sample taken about $i$:
\begin{equation}L_{j}(x_{i})=\left.\frac{\mbox{Det}[v_{j}]}{\mbox{Det}[v]}\right|_{i}\end{equation}
where the subscripts indicate that the Vandermonde matrix defined at pixel $j$ has the column $i$ replaced by a column of undetermined positions $x$, i.e., for the one-dimensional case

\begin{figure}[H]\centerline{\includegraphics[height=90mm,width=180mm]{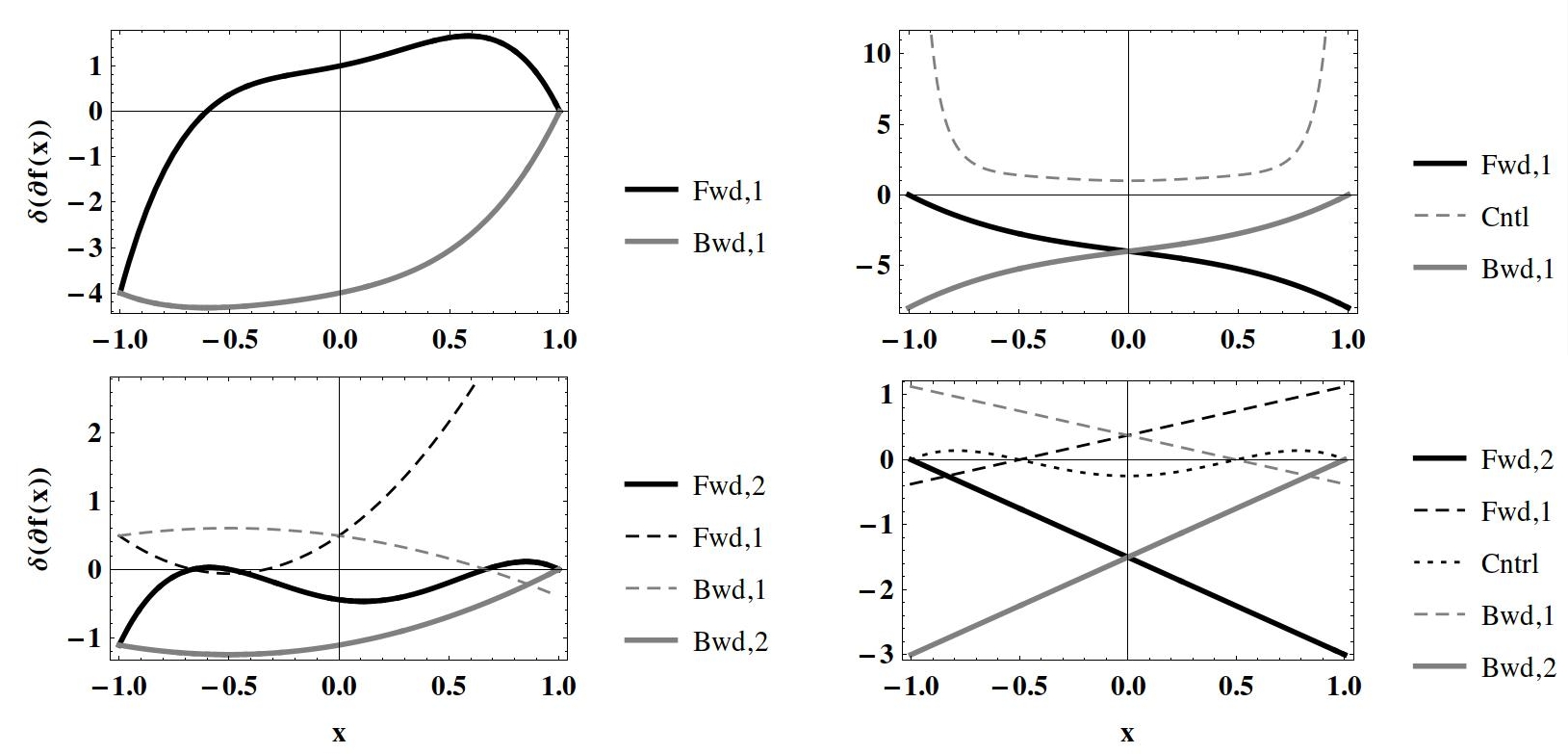}}\caption[]{Accuracy \begin{math}\delta (\partial f(x))=\partial f_{\mathrm{exact}}(x)-\partial f_{\mathrm{fin.diff.}}(x)\end{math} of (clockwise from top-left) 2-, 3-, 5-, and 4-point finite-difference schemes on 1-d grids. The 2- and 4-point schemes are calculated with a varying stencil leftmost pixel location, whilst the 3- and 5- point schemes feature a varying central pixel location. All the  remaining pixels are fixed. We use the test function \begin{math}f(x)=x^{5}\end{math}, since there are fewer datapoints available in these schemes than can specify the function exactly. The differences blow up only as the coordinate variation produces degeneracy in the datapoints. In the key, forward-, backward-, and central-differences are abbreviated "fwd", "bwd", "cntrl" respectively, with the following number specifying the distance (in points) of the focal point from the grid centre.}\end{figure}

\begin{equation}v_{2}=\left(\begin{array}{ccccc}x_{1}^{0}&x^{0}&x_{3}^{0}& \cdots & x_{n}^{0} \\ \vdots & \vdots & \vdots & \ddots &\vdots 
\\ \vdots & \vdots & \vdots & \ddots & \vdots
\\ x_{1}^{n-1}&x^{n-1} & x_{3}^{n-1} & \cdots & x_{n}^{n-1}\end{array}\right),\end{equation}
in order to find the interpolated function value at a given pixel $i$ (not necessarily part of the pre-sampled pixel distribution), with $x$ corresponding to the positions $x_{j}$ in Eq. (17), placed in column $i$ in $v_{i}$. The Vandermonde array is then the one-dimensional case of a more general geometric array, related to interpolation over $d$ dimensions. The specification of the geometric array is dictated by the dimensionality of the space being interpolated over and by inspection of the geometry (and thus possible derivatives) of the pixel sample one is working with; for example, a square grid of $4\times 4$ points in ($x$, $y$) allows a polynomial specified by powers of up to $x^{3}$, $y^{3}$ and combinations thereof; the geometric array would then include combinations of powers of $x$, $y$ up to such a limit. For the purpose of performing derivatives, we replace $x_{i}\rightarrow\Delta_{ij}=x_{j}-x_{i}$.

\section{The HEALPix Sphere}

\noindent HEALPix [6], a \textit{Hierarchical Equal Area isoLatitude
Pixelisation} scheme for the sphere, is the most widely used software for construction and analysis of full-sky CMB maps. The lowest-resolution partitioning is into 12 equal area pixels. Each pixel is assigned a unique identification number (in either a ``nested'' or ``ringed'' numbering scheme), and is surrounded by 8 other pixels, except in the polar cap where some of the pixels are surrounded by 6 or 7. The map resolution is specified by the parameter 
\begin{equation}N_{\mathrm{side}}=2^{N_{\mathrm{order}}},\indent N_{\mathrm{order}}\in|\mathbb{Z}|\end{equation}
with each map of $12N_{\mathrm{side}}^{2}$ pixels composed of \begin{math}4N_{\mathrm{side}}-1\end{math} isolatitude rings; at each level higher in resolution, the pixels are subdivided into 4 equal area pixels in the higher-resolution map. The rings immediately by either pole consist of 4 pixels (independent of resolution), increasing by 4 pixels per ring for each ring increment toward the equator, up to a maximum \begin{math}4N_{\mathrm{side}}\end{math} pixels in the equatorial rings.
\\The HEALPix sphere is an interesting case study, since it is a semi-regular distribution on a coordinate system with a pathology: the multi-valuedness of \begin{math}\phi\end{math} at the poles --- hence differences in ($\theta, \phi$) across the pole are ill-defined (see Appendix B for further details). It does not matter if we do not directly sample the pole point itself --- merely crossing the pole with a differencing stencil is enough to complicate the calculations, since the interpolating polynomial covers the whole region. Similarly, we must be careful at the \begin{math}\phi=0/2\pi\end{math} boundary. However, at such a boundary here we can simply reassign the differences to the smallest of the two possible paths across the sphere between the points we are differencing.
\\To test the software, the derivatives of harmonic functions on the sphere are computed, since primordial cosmological point signals are not expected. This implementation uses LAPACK and the truncated SVD technique [17, 18] for ill-conditioned matrices (in the limit of large stencils and irregular geometries). The geometry for each pixel stencil taken is given by a square-geometry differenced geometric array
\begin{equation}v'=\left(\begin{array}{cccc}1 & \cdots & \cdots& 1 \\ \Delta \theta_{1} & \cdots & \cdots & \Delta \theta_{n} \\ \Delta \phi_{1} & \cdots & \cdots & \Delta \phi_{n} \\ \Delta \theta_{1}\Delta \phi_{1} & \cdots & \cdots & \Delta \theta_{n}\Delta \phi_{n} \\ \vdots & \cdots & \cdots & \vdots \\ \Delta \theta_{1}^{n-1}\Delta \phi_{1}^{n-1} & \cdots& \cdots & \Delta \theta_{n}^{n-1}\Delta \phi_{n}^{n-1}\end{array}\right),\end{equation}
where we have chosen to work in the ($\theta$, $\phi$) basis for computational convenience (one does not necessarily expect the Stokes fields to be a polynomial). This choice is determined by issues arising from analysis at the pole; whilst the analytic operators we are calculating are necessarily covariant, discretization complicates matters at the pole. The utilised alternative is  choosing the pixel stencils in $(\theta,\phi)$ such that they do not cross the pole at all, but instead progress toward an outer-difference scheme around the pole.
\\In the particular case of extracting $E$- and $B$-modes, the pole also presents other numerical problems; ways to deal with these problems include ``rotated oversampling'' around the pole, rotating the polar data to the equator and rotating the resulting fields back, or by simply discarding the polar pixels. These will be discussed briefly after the next subsection.

\subsection{Structure of the Algorithm}
The algorithm is as follows (see Fig. \ref{fig:sampling1} for some visual examples):
\begin{itemize}
\item Construct an approximately square stencil of nearest-neighbour pixels, one for each pixel. This is achieved by taking the unique set of nearest neighbour pixels for the focal pixel, and then repeating recursively for the neighbour pixels until the specified stencil radius is satisfied. If the array cannot be filled with corresponding pixel numbers (such as in the case that a pixel has less than 8 neighbouring pixel), then the remaining elements are assigned an identification value of -1. Note that since the HEALPix nearest-neighbours-finding routine always calls the neighbour pixels in the same geographic order, then pixels with identical surrounding geometries will necessarily have identical stencil arrays. 
In other words, the code is structured in such a way that the symmetries of the HEALPix grid are preserved also in the stack of pixel stencils taken. Our notation for the stencils is as follows: a stencil of order \begin{math}n\end{math} contains at most (existence and masking-dependent) precisely \begin{math}(n+1)^{2}\end{math} pixels. Since we wish to bias toward central differences, each \textit{n} is even, corresponding to a radius of \begin{math}n/2\end{math} pixels about the central focus pixel. The stencil order \textit{n} is initially the same for all pixels.

\begin{figure}[H]\centerline{\includegraphics[height=80mm]{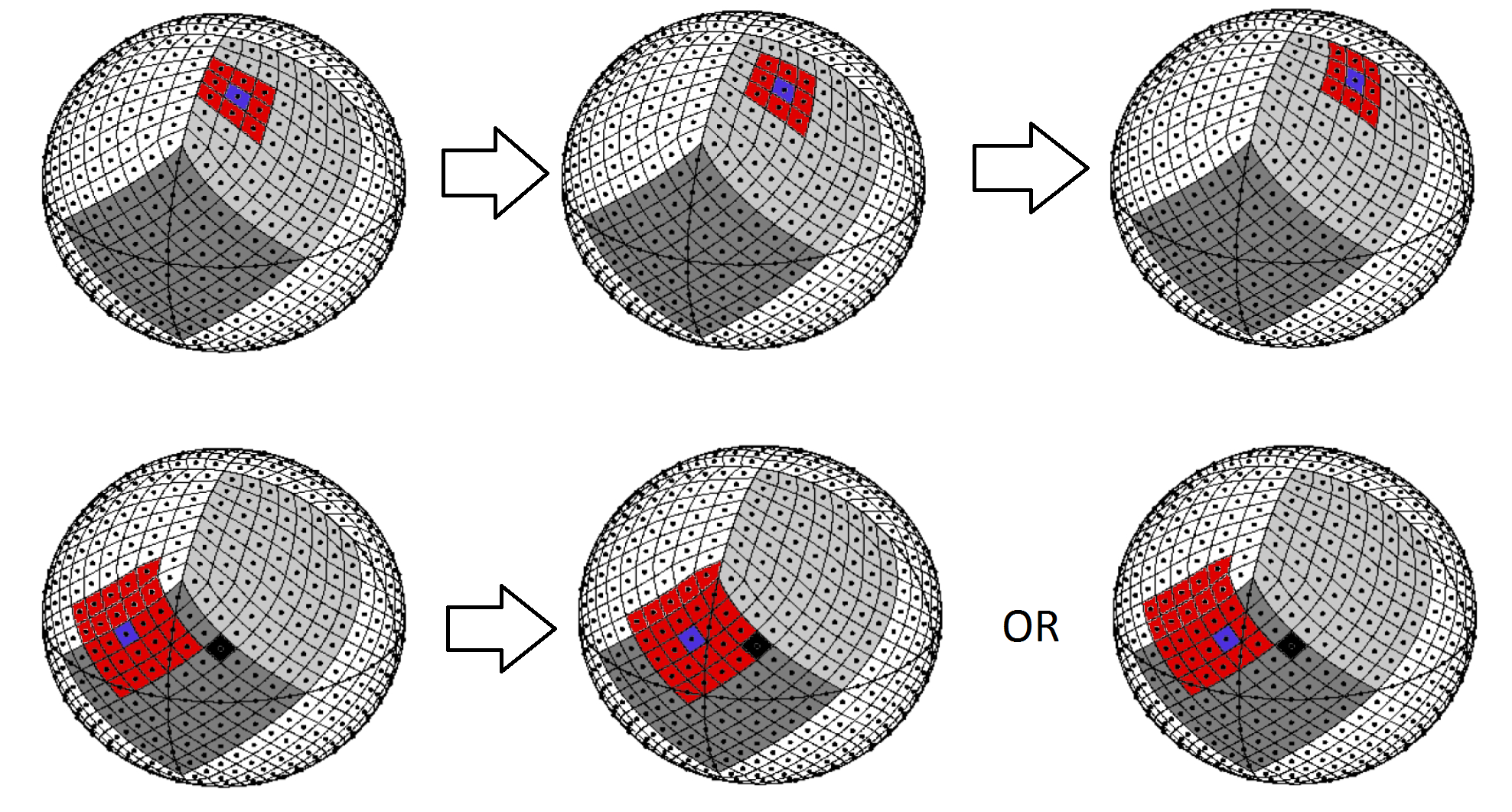}}\caption[Standard MasQU sampling method]{Standard sampling method, across orders, across the sphere. Top row: iteration of the $n=2$ calculation across a ring on the maskless sphere. The coloured pixels populate the pixel stencil whose geometry is used in calculating the pixel weights, which are used for approximating the value of a derivative at the blue focal pixel. Bottom row: for a sphere with a masked pixel (in black), an example is made of the $n=4$ calculation. The left diagram shows the scheme when the masked pixel is not within the pixel stencil. The centre and right diagrams are central and outer-difference schemes; the outer-differenced scheme contains more information than the central-differenced scheme in the case shown.}\label{fig:sampling1}\end{figure}

\item Perform ``re-mapping'' for pixels $p_{i}$ surrounded by one or more masked pixels --- search amongst the stencils of the surrounding pixels $p_{j}$ for the ``best'' stencil (the stencil with the most available pixels and closest to the central focus). A weight system of
\begin{equation*}W=\left\{\begin{array}{cc}0&\mbox{If pixel is missing or in a cut region}\\1&\mbox{Otherwise}\end{array}\right.
\end{equation*}
is applied.
Pixels without enough information in their stencil to perform the requisite derivatives (three available pixels with a unique position $\theta$, and the same for $\phi$ ) are discarded.
\item Analyse the mapped difference geometries to single out a smaller number of unique stencil geometries --- these characterize a unique set of stencils which provide all the necessary weights for calculation across the sphere, saving computation time. In the unmasked case, there are also symmetries between the north and south polar regions and also the quarters of each hemisphere which can be taken advantage of to cut down computation time.
\item Find solutions to the linear equation
\begin{equation}v'w=\delta,\end{equation}
whose 1-d analogue is Eq. (16), corresponding to each of the derivatives required --- each set of weights is unique for a given stencil geometry and derivative order (except in trivial cases). The geometric array is by default that for a regular square array. In the case that there is not a full square stencil available, we remove rows (starting from the lowest row/largest difference powers) and the corresponding columns of the geometric array such that the array remains square. It is probably more rigorous to remove rows using geometric considerations instead (i.e., an analysis of which derivatives can be calculated from the stencil), but this has yet to be implemented and does not significantly affect the rest of the analysis since only the derivatives up to second order in ($\theta$, $\phi$) are calculated. Furthermore, the additional time cost for such a procedure might not be a good trade-off. Solutions to the linear equations are found using LAPACK and the QR decomposition or SVD technique, depending on the pixel sample.
\item Repeat until all the calculations for each pixel are finished, and form the bi-Laplacians. Compute the pseudo-$C_l$ power spectra of the resulting $\nabla^{4}e$ and $\nabla^{4}b$ maps and remove the power contributed by the bi-Laplacian operator.
\item In the presence of a mask, one should apodize the signal since masking redistributes signal power. An apodization subroutine will be made available with the software, whilst there is an optimal method for CMB studies due to Smith \& Zaldarriaga [19]; we leave the analysis of this to a later paper.
\item Output the derivative maps, the power spectra, and the weights used (these can be recycled once calculated for the first time) in FITS format.

\end{itemize}

\noindent The above methods can be modified in the software for a given geometric scheme (be it for alternative pixellization schemes on $\mathbb{S}^{2}$ such as GLESP, or non-spherical schemes) and a given desired derivative. Higher-dimensional generalizations have not been implemented but would be trivial. 
Testing the software for accuracy follows the process of taking pre-specified $E$- and $B$-mode harmonic coefficients, calculating the full-sky $Q$ and $U$ maps from these coefficients, operating  on the sky maps to produce the bi-Laplacians, and then converting the power spectra of the $\nabla^{4}e$ and $\nabla^{4}b$ fields to that of the $E$- and $B$-modes for comparison with the original input spectra:
\begin{equation*}a_{lm}^{E,B}\rightarrow C_{l}^{E,B}\rightarrow Q,U\overbrace{\rightarrow}^{\partial Q, \partial U}\nabla^{4}e,\nabla^{4}b\rightarrow C_{l}^{E,B}\end{equation*}
where we also compare the full scalar field maps $\nabla^{4}e$, $\nabla^{4}b$ with the sum of the original harmonic coefficients as in Eq. (2). The accuracy of the software, as operated on a test harmonic function, is shown in Fig. 3.

\begin{figure}\centerline{\includegraphics[height=158mm]{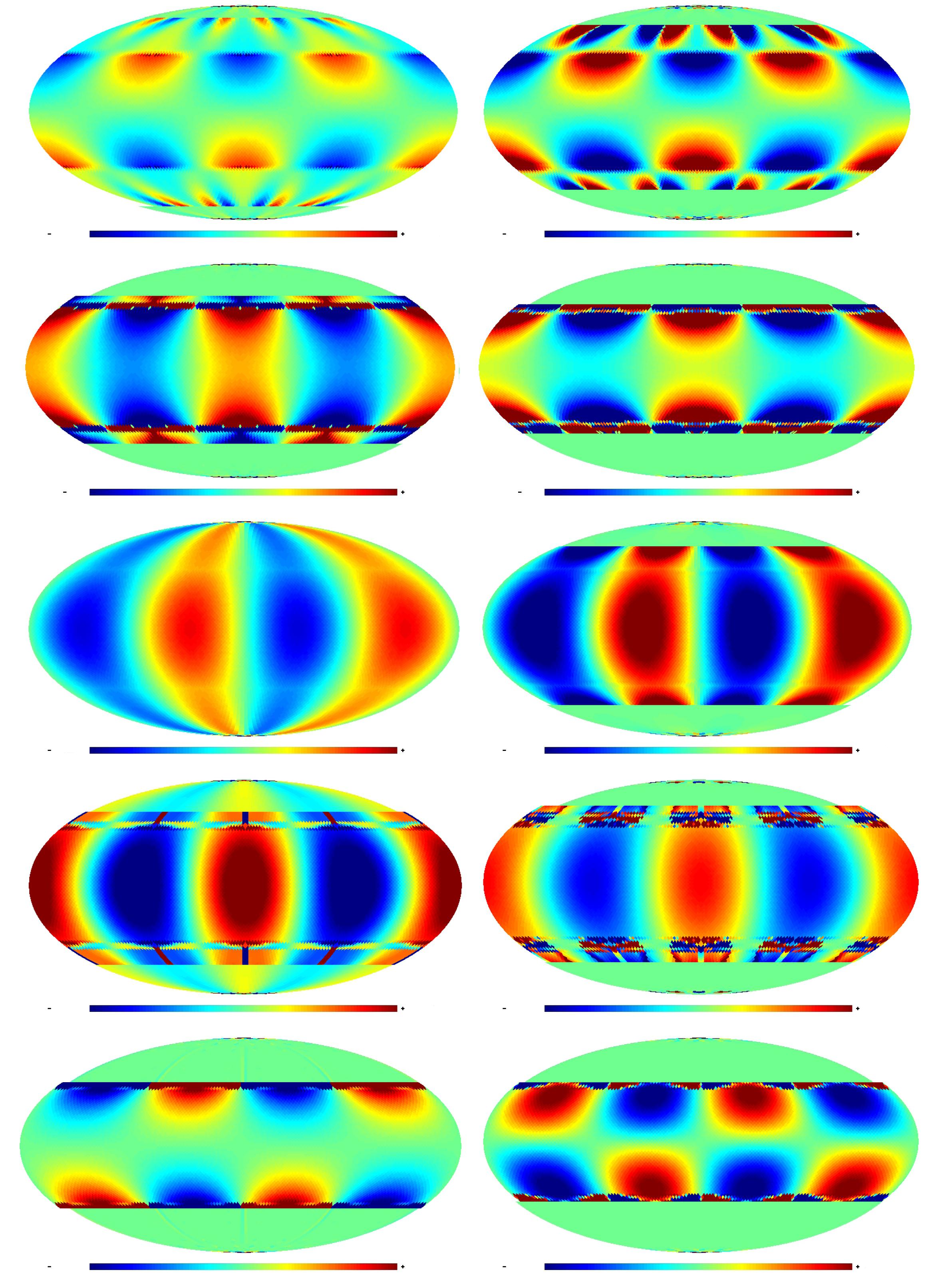}}\caption[]{$n=2$ and $n=4$ accuracy (across rows) at $N_{\mathrm{side}}=128$ for derivatives $\partial_{\theta},\partial_{\theta\theta},\partial_{\phi},\partial_{\phi\phi},\partial_{\theta\phi}$ (down columns) of $Y_{22}(\theta,\phi)$. The absolute maximum values for each image in the figure array are $\sim\left(\begin{array}{ccccc}1.5\times10^{-3}&0.1&4.5\times10^{-4}&2.6\times10^{-3}&3.1\times10^{-2}\\1.5\times10^{-2}&0.8&1.5\times10^{-3}&3.1\times10^{-3}&9.0\times10^{-2}\end{array}\right)^{T}$. Since the error in the polar cap is much larger than the equatorial region error for all maps bar $\partial_{\phi}$, the values in the equatorial regions in these maps has been limited to $\left(\begin{array}{ccccc}\pm10^{-6} & \pm2\times10^{-6} &\pm2\times10^{-6} &\pm2\times10^{-6} &\pm10^{-6} \\ \pm10^{-3} &\pm10^{-3} &\mathrm{n/a} &\pm10^{-3} &\pm2\times10^{-4} \end{array}\right)^{T}$, resulting in the
observed discontinuous 'bands' at the polar cap boundaries. The figure shows that the accuracy improves with stencil size, and is worst at the pole (due to a combination outer-differencing error (for $\partial_{\theta}$ and $\partial_{\theta\theta}$), and increased differencing values for the $\Delta$ terms). One should not simply increase the stencil order to improve accuracy, since: 1) The time complexity increases; 2) We are limited to a maximum stencil size of $(2N_{\mathrm{side}}+1)^{2}$. (3) The example ($l$, $m$) given has a relatively small magnitude pole problem error for harmonic functions (See Fig. 5). Thus the convergence rate with stencil size is too slow.}\end{figure}

\noindent The SVD technique separates a matrix A into 
\begin{equation}A=\alpha^{T}\Sigma \beta\end{equation} (where \begin{math}\Sigma\end{math} is the diagonal array of singular values) and is known for yielding optimal linear solutions in the presence of near-singular matrices. In the near-singular case, the smaller singular values can be dominated by round-off error leading to dramatic errors in the solution. For large geometric arrays, the bottom-row elements are most likely to suffer from round-off error since they are large powers of small numbers. This means the singular value array must be truncated below some numerical threshold in order to yield reasonable numerical results, and the corresponding inverted elements in the inverted singular value array replaced with zero; truncation error will contribute to any inaccuracy of the calculation. Since the different resolutions correspond to a scaling of geometries on the HEALPix sphere, the threshold should be defined by the ratio $\Sigma_{ii}/\Sigma_{11}$. Optimal truncation may depend on some non-machine aspects: array size and geometry. Whilst this problem has not been solved generally, empirically-determined truncation thresholds for the unmasked sphere as-a-whole (i.e., individual stencil geometries have not been studied), which give the best global calculation accuracy, have been implemented for the first few stencil orders. Masking has not been fully SVD-optimised (we use the same thresholds as for unmasked maps) since this study requires generalization to any number of pixels in a range of distributions.
\\Finally, we consider time complexity. An approximation of the complexity of the algorithm on the unmasked sphere can be given by $\sim N_{\mathrm{geom}}\mathcal{O}\left((n+1)^6\right)$, since $n$ is the order of the calculation, there are $(n+1)^2$ elements in the geometric array, and matrix inversion for a $k$-by-$k$ matrix is $\sim\mathcal{O}(n^3)$. $N_{\mathrm{geom}}$ is the number of unique stencil geometries on the HEALPix sphere, which cannot be more than $N_{\mathrm{pix}}/8$ due to basic hemispherical and quadrant symmetries of the HEALPix grid. In the maskless case, taking only one hemisphere and quadrant, the unique stencils are: $N_{\mathrm{pix-in-ring}}/4$ for each of the polar rings, yielding $\sum^{N_{\mathrm{side}}}_i i$ for each polar region; the first $n$ rings beyond the polar boundaries have $N_{\mathrm{side}}$ unique stencils each; and the remaining $(N_{\mathrm{side}}-n)$ have one unique stencil each, yielding
\begin{equation}N_{\mathrm{geom}}=\frac{N_{\mathrm{side}}(N_{\mathrm{side}}+1)}{2}+n(N_{\mathrm{side}}-1)+N_{\mathrm{side}}.\end{equation}
While it can be seen that a first-time implementation of the code is out-performed by the standard HEALPix methods (\textit{anafast}) in terms of speed, one needs calculate the weights only once. Fig. 4 compares standard raw full-sky MasQU and HEALPix computation times on a low-end machine; whilst HEALPix is an order of magnitude quicker than the first-time MasQU calculation, using pre-calculated weights makes the MasQU computation easily competitive with HEALPix.

\begin{figure}[ht]\centerline{\includegraphics[height=80mm]{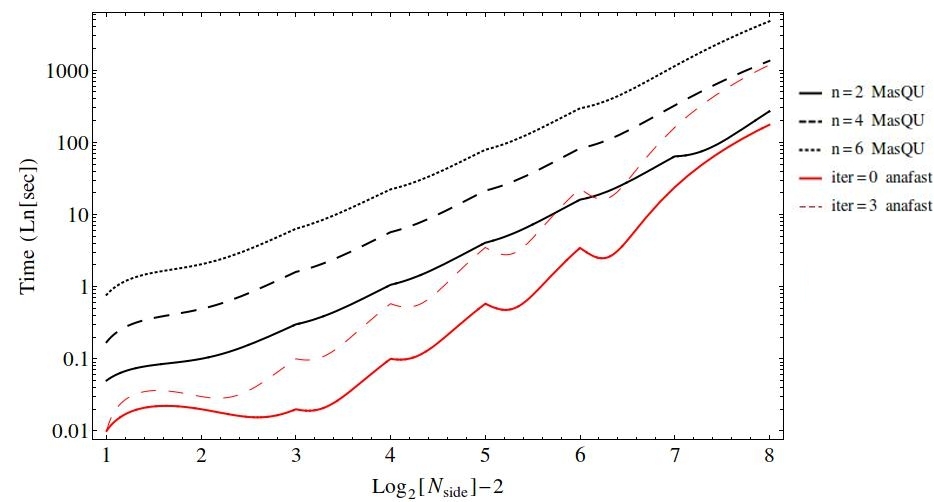}}\caption[]{Computation time $t$ for maskless MasQU and HEALPix calculations (for \textit{anafast} iteration $iter$), across $N_{\mathrm{side}}$. Masked calculations with sky coverage $f_{\mathrm{sky}}$ scale as $f_{\mathrm{sky}}t_{\mathrm{fullsky}}$; for limited sky coverage, the calculation times can become competitive with HEALPix on the same machine.}\end{figure}

\subsection{The ``Pole Problem'' and Rotated Sampling}

For polar pixels, the algorithm is amended as follows:
\begin{itemize}
\item Define polar pixel ``re-mappings''; for each pole-crossing pixel (referred to as $p_{\mathrm{pol}}$), we search among its stencil pixels for the set of pixels whose stencils do not cross the pole. We then select from this set the subset of pixels that satisfies the requirements that minimize their distance on the sphere from $p_{\mathrm{pol}}$ in order that the new calculation scheme is as close to a central-difference scheme as possible, and from this subset choose the pixel which is closest to the pole (referred to as $p_{\mathrm{npol}}$). We then reassign the stencil of $p_{\mathrm{pol}}$ from a central-difference in the $p_{\mathrm{pol}}$ stencil basis to an outer-difference in the chosen $p_{\mathrm{npol}}$ basis.
\end{itemize}

\begin{figure}[H]\centerline{\includegraphics[height=80mm]{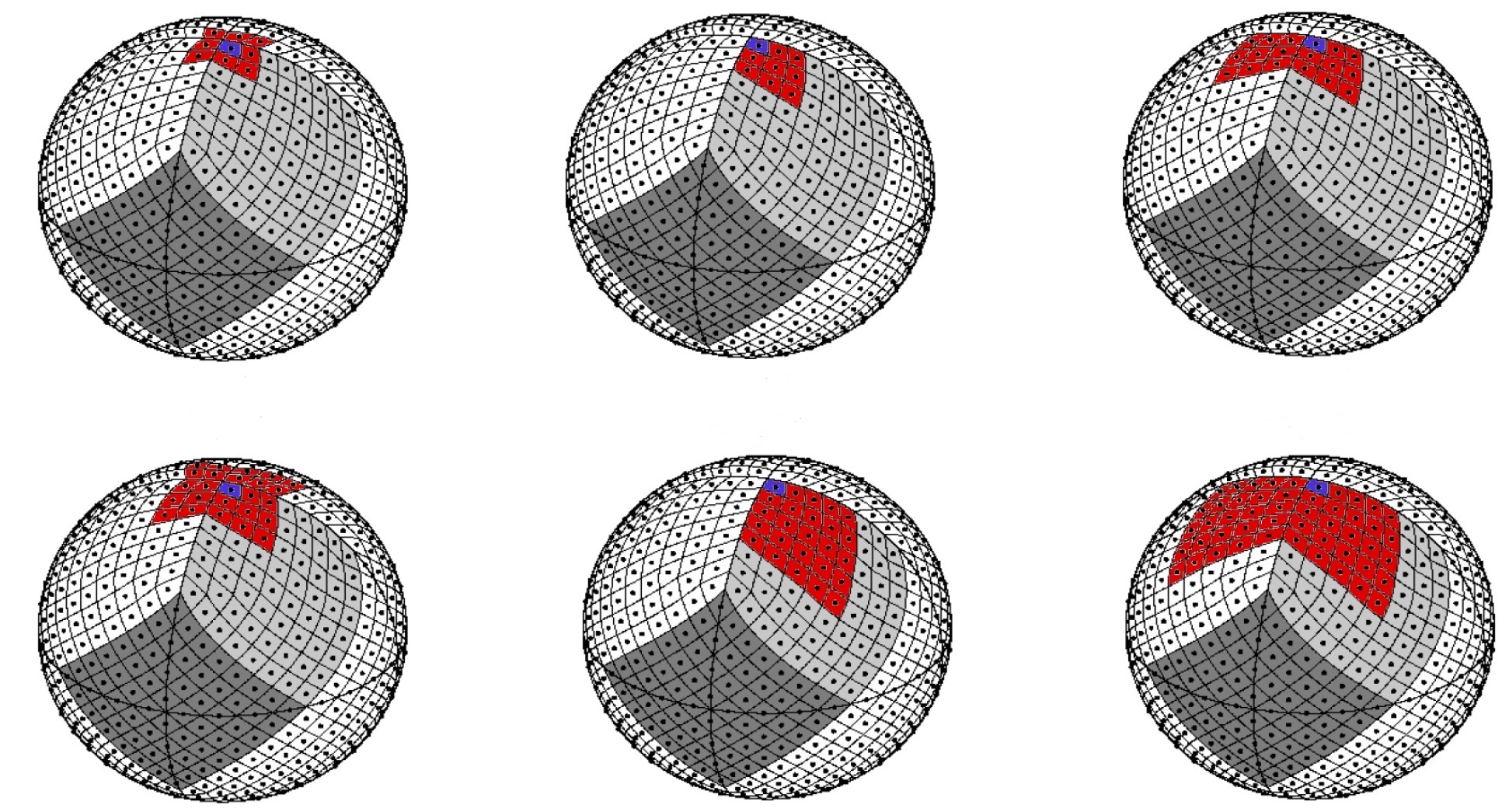}}\caption[MasQU polar sampling]{Various polar sampling methods. Left diagram: standard pole-crossing stencils. Middle: outer-differenced stencils. Right diagram: `Doubled' sampling, equivalent to $\phi$-rotated sampling with $\delta\phi=\pi/2$. The top and bottom rows correspond to $n=2$ and $n=4$ sampling respectively.}\end{figure}

\noindent The full array of pole problems for schemes not crossing the pole are as follows. Firstly, we can see from our one-dimensional tests  (Fig. 1) that while the forward- and backward-difference schemes are of the same magnitude error as the central schemes, they generally perform slightly worse than equivalent central difference schemes. The same is true for comparisons of higher- to lower-order derivatives. This is because higher-order derivatives require more sampling pixels, and hence a larger basic stencil size, than lower orders. We can also see this in the HEALPix scheme for derivatives of the $s=0$ spherical harmonics (Fig. 3), hence the higher errors at the poles. We refer to this as ``outer-differencing'' error.
\\Secondly, the $\Delta\phi$ values between sampled pixels increase toward the pole due to the lower pixel sampling per HEALPix ring toward the pole. At its most extreme, the pixel rings immediately surrounding either pole have only 4 pixels each, with a separation \begin{math}\pi/2\end{math}. Since the errors in our differencing scheme depend on powers of \begin{math}\Delta\phi\end{math}, then we expect a lower accuracy around the pole. This we refer to as the ``polar \begin{math}\Delta\phi\end{math}'' problem. It is apparent in all cases that (i) second-order derivatives perform slightly worse than first-order derivatives; (ii) $\partial_{\theta\theta}$ performs worst. This is no surprise given that the polar cap pixels are only central in \begin{math}\phi\end{math}; (iii) large-magnitude functional variation in \begin{math}\phi\end{math} about the pole seems to correlate with large polar errors (Fig. 6).

\noindent We can look first at ameliorating the outer-difference and polar $\Delta\phi$ error. Fig. 7 includes calculations when the stencil size is increased, but only around the pole.
\\A third, more drastic problem comes from the construction of the terms we wish to calculate. In the continuum limit, these are
\begin{equation}\begin{split}\nabla^{4}e=-(-2-\csc^{2}\theta\partial_{\phi\phi}+3\cot\theta\partial_{\theta}+\partial_{\theta\theta})Q-2\csc\theta(\cot\theta\partial_{\phi}+\partial_{\theta\phi})U
\\ \nabla^{4}b=2\csc\theta(\cot\theta\partial_{\phi}+\partial_{\theta\phi})Q-(-2-\csc^{2}\theta\partial_{\phi\phi}+3\cot\theta\partial_{\theta}+\partial_{\theta\theta})U.\hspace{0.3cm}\end{split}\end{equation}

\noindent This is quite a delicate combination due to a number of $\csc\theta$ and $\cot\theta$ terms, which clearly blow up as we approach the pole. The contributor of the largest error is then the $\csc^{2}\theta\partial_{\phi\phi}$ term, which will blow up any errors in the discrete approximation to \begin{math}\partial_{\phi\phi}\end{math}. This is referred to as the ``blow-up'' problem, and accentuates the first two problems.

\begin{figure}[H]\centerline{\includegraphics[height=90mm]{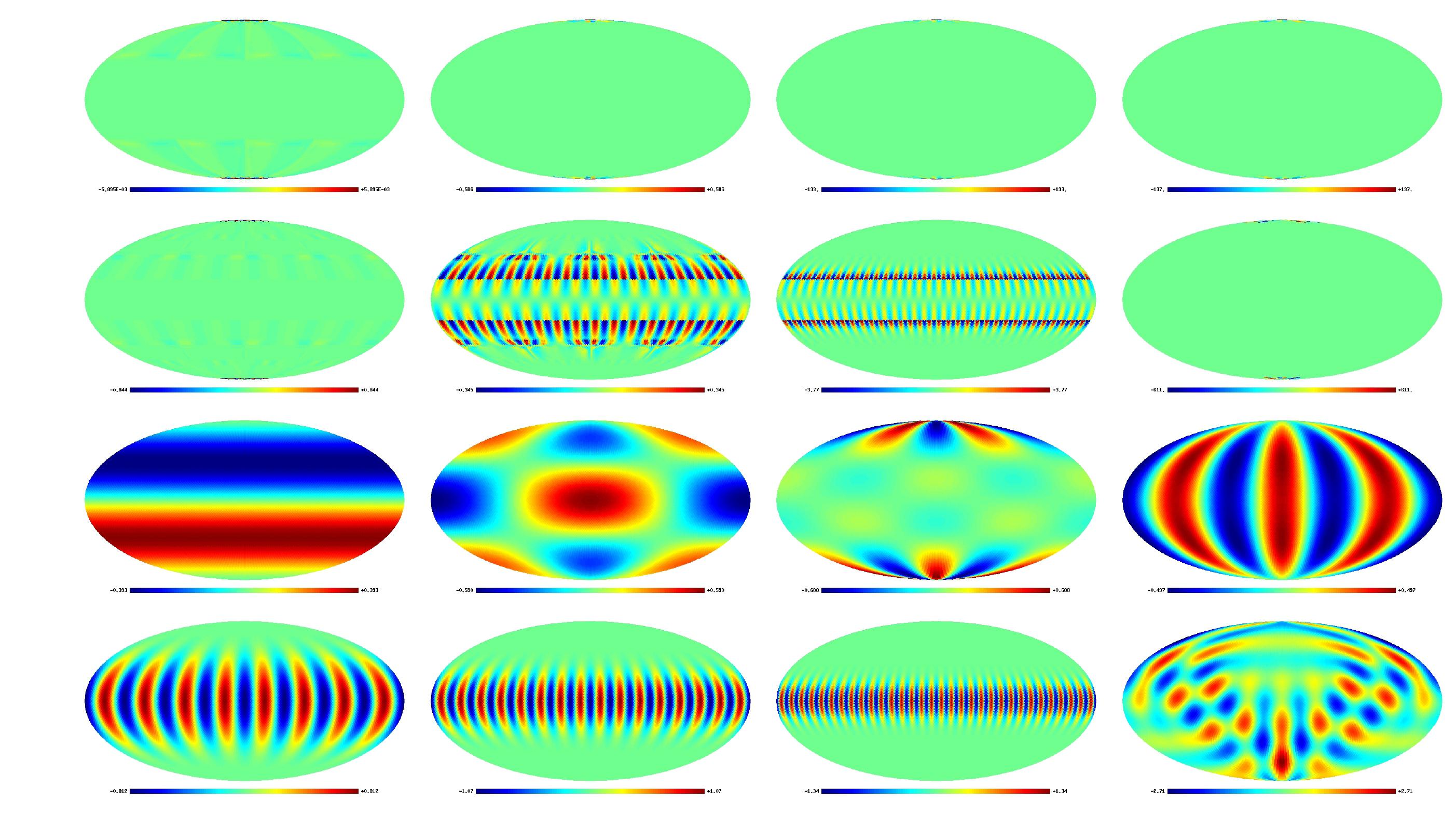}}\caption[]{Top 2 rows: $n=4$ accuracy maps of $\nabla^{4}b$-fields with $N_{\mathrm{side}}=32$ for a range of unit-$a^{E}_{lm}$ sources; the maps are symmetric in errors, with maximum absolute values of $\left(\begin{array}{cccc}
5.9\times10^{-3}&0.6&133&137 
\\0.8&0.3&3.8&611\end{array}\right)$.  
Bottom 2 rows: $Q$ maps corresponding to the sources given: \\$a_{lm}^{E}=\left(\begin{array}{cccc}
\delta_{l3}\delta_{m0}&\delta_{l3}\delta_{m1}&\delta_{l3}\delta_{m2}&\delta_{l3}\delta_{m3} 
\\ \delta_{l8}\delta_{m8}&\delta_{l16}\delta_{m16}&\delta_{l32}\delta_{m32}&\sum_{l'=0}^{9}\sum_{m'}\delta_{ll'}\delta_{mm'}\end{array}\right)$. Whilst the absolute maximum values are of order $\sim1$ for each $Q$ map, the polar errors peak for $m=2$ and $m=3$, and the magnitude of the polar errors at fixed \textit{m} increases with \textit{l}. ($l$, $m$) = ($32$, $32$) maps have been included to show how error can propagate at the equator, due to variation in \begin{math}\phi\end{math} greater than that modellable by the stencil.}\end{figure}

\noindent The combination of these effects leads to dramatic problems at the pole for realistic (CMBFAST-generated [20]) sky maps: in Fig. 8 we can see the operation on realistic $B$-mode-free maps. Even for a large stencil the error swamps the real-space $\nabla^{4}e$ signal; for a non-zero $\nabla^{4}b$ signal the problem is then more drastic due to the $\nabla^{4}b$ signal's small magnitude.
\\Taking derivatives using larger stencils about the pole is not always advantageous. Instead, a method for dealing with the pole is by rotated sampling; rotate the sphere by some small amount \begin{math}\delta\phi<\Delta\phi\end{math} and perform derivatives on a ``doubled'' stencil. Fig. 7 shows that the error is greatly reduced in this manner. There is though the caveat that it is necessary to sample the harmonic coefficients accurately in order to effectively remove the pole problem. The polar \begin{math}\Delta\phi\end{math} problem ensures that taking ``discrete'' doubled sampling (calculating over a stencil which includes the next-neighbour-in-\begin{math}\phi\end{math}'s stencil) is not nearly effective enough. Of course, for the rotated sampling case we can always perform the rotation more than once in order to improve the accuracy. In the case of discrete doubling, we are limited to using  \begin{math}N_{\mathrm{pix-in-ring}}/2\end{math} adjacent stencils, so as not to have a pole-crossing calculation.
\\Another alternative is to rotate the underlying coordinate system by $\delta\theta=\pi/2$ which rotates the pole to the equator of a new coordinate system. This allows us to calculate the rotated scalar and pseudo-scalar field pole pixels using a central-difference scheme whilst avoiding the pole issues of the $\csc\theta$-type terms. Rotating the resulting calculations back to the pole results in ringing from the pixel boundary, which can be tempered by calculating a larger region of rotated pixels since ringing dies down further from a discontinuous boundary, or by applying a tapering function; rotating to the equator is recommended if the polar pixels are needed.
\\In the absence of a both quick and accurate method for dealing with the pole, one is reduced to simply removing the offending pixel rings; for most calculations this is recommended since the polar pixel region is negligible for large (i.e., \textit{Planck}-type) maps. A comparison of various oversampling techniques is seen in Fig. 7.

\begin{figure}[H]\centerline{\includegraphics[height=100mm]{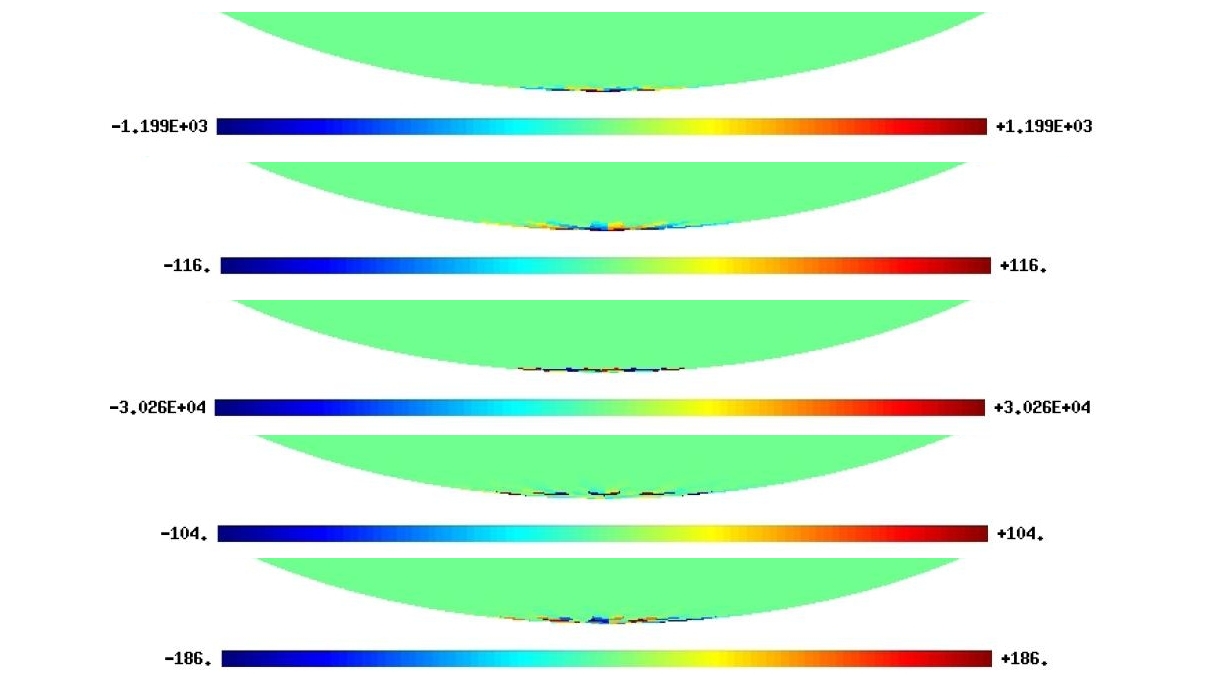}}\caption[]{$n=2$ $\nabla^{4}b$-fields, with $N_{\mathrm{side}}=32$: various oversampling techniques; only the polar values are of interest here. From top to bottom: (i) Original results for the source $a^{E}_{lm}=\delta_{l2}\delta_{m2}$, $a_{lm}^{B}=0$, (ii) with an $n=4$ stencil about the pole, (iii) \begin{math}\delta\phi=0.01\end{math} HEALPix-reconstructed ($l_{\mathrm{max}}=2N_{\mathrm{side}}$)rotated doubled sampling, (iv) $\delta\phi=0.01$ analytic rotated doubled sampling, (v) ``discrete'' doubled sampling. The ``discrete'' doubled sampling mode performs better than the standard $n=2$ but falls short of the rotated sampling; with an accurate map reconstruction method there is nothing in principle to forbid an $n$-tupled rotation sampling, which would bring the pole error down quickly.}\end{figure}

\begin{figure}[ht]\centerline{\includegraphics[height=70mm]{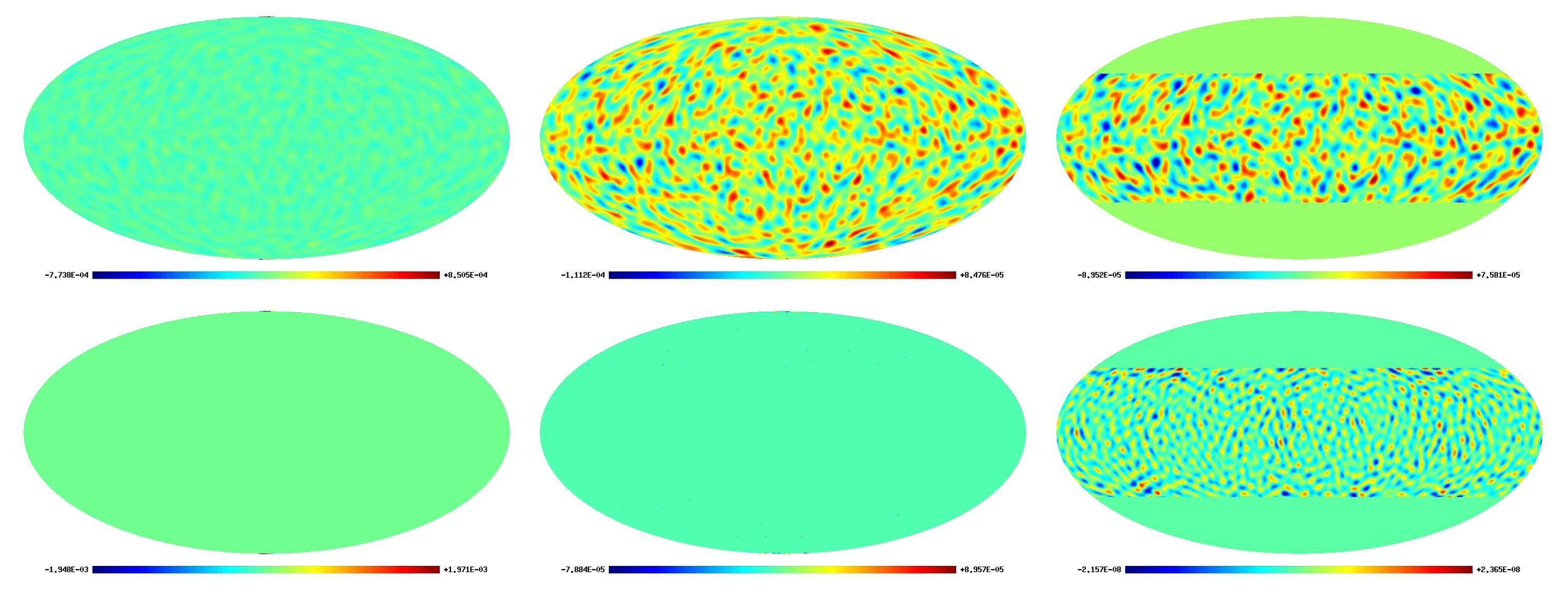}}\caption[]{Top row: $\nabla^{4}b$ map ($n=4$) generated from CMBFAST $Q$ and $U$ maps, with tensors. Bottom row: $\nabla^{4}b$ map ($n=4$) generated from CMBFAST $Q$ and $U$ maps, no tensors. The map resolution is $N_{\mathrm{side}}=128$, and all the other parameters are the defaults from the LAMBDA [21] online tool page given in Table I. Columns, left-to-right: No ring removal, 1st north \& south polar rings removed, equatorial region only. In the no-tensors scenario, at the power spectrum level the pole problem could provide a false-positive detection; at the level of the map, we can distinguish these by eye --- inconsistency between polar and equatorial regions in a given map, as seen in the middle and right-hand images, is a result of the differencing error dominating the calculated map. This then amounts to a consistency criterion by which to check for a false positive detection. The irregular geometries outside the equator induce a larger error than at the equatorial region. For the equator, the cut-off length is defined by the ring at which the stencil geometries become irregular, and is hence order \begin{math}n\end{math}-dependent.}\end{figure}

\section{Masking \& noise performance}

\subsection{Estimators}

\noindent For cosmology, it is necessary to calculate the errors on these terms in harmonic space such that one can define the accuracy of the power spectra estimator. For example on a flat 1-d space we can compute the estimator for the transform of a $1^{\mathrm{st}}$-order derivative, 
\begin{equation}F(x)=\sum_{k}\hat{f}_{k}e^{ikx}\Rightarrow\partial_{x} F=\sum_{k}ik\hat{f}_{k}e^{ikx}\end{equation}
where $\hat{f}_{k}$ is the Fourier power in $F$ for mode $k$. On a discrete grid there is instead the 1-point forward difference equation
\begin{equation}\tilde{\partial}_{x}F=\sum_{k}\hat{f}_{k}\frac{i\sin(k\Delta)e^{ikx}}{\Delta}\end{equation}
where the $\sim$ over the derivative operator denotes a discrete derivative, hence the finite-differenced spectrum $\tilde{\mathcal{P}}_{k}$ always underestimates the true spectrum $\mathcal{P}_{k}$:  
\begin{equation}\tilde{\mathcal{P}}_{k}=\mathrm{sinc}^{2}(k\Delta)\mathcal{P}_{k}.\end{equation}
In the following, we define an \textit{n-point stencil} as a collection of $n$ pixels over which the finite-difference calculation in Eq. (15) for a given point on a grid is made. The full flat-sky operators on a regular grid with 3-point stencils are then given by
\begin{equation}\begin{split}\tilde{D}^{+}_{0}F=-\sum_{k}\frac{2}{\Delta^{2}}(\cos( k_{2}\Delta)-\cos(k_{1}\Delta))\hat{f}_{k}e^{i\mathbf{k}\cdot\mathbf{n}}
\\ \tilde{D}^{-}_{0}F=-\sum_{k}\frac{2}{\Delta^{2}}(\sin(k_{1}\Delta)\sin(k_{2}\Delta))\hat{f}_{k}e^{i\mathbf{k}\cdot\mathbf{n}}.\hspace{4mm}\end{split}\end{equation}
For more general 1-d derivatives, we have to start from
\begin{equation}\tilde{\partial}_{x}F_{j'}=\sum_{k}\hat{f}_{k}e^{ikx}\sum_{j} w_{j}e^{ik\Delta_{j,j'}}\end{equation}
and induce simplifications at this point using grid symmetries. There is then no simple sinusoidal statement for general irregular grids but the estimator is not difficult to calculate. The first-derivative spectral estimator is then proportional to
\begin{equation}\tilde{\mathcal{P}}_{k}\propto\left(\sum_{j} w_{j}e^{ik\Delta_{j,j'}}\right)^{2}\end{equation}
where the full irregular flat sky discrete operators are found using the general solution given by Eq. (A28) in Appendix A
\begin{equation}w_{ij}^{(m)}=\frac{(\partial_{\Delta})^{m}\left[\prod_{l=1,p_{l}\neq i}^{n}\Delta_{p_{1}j}\cdots\Delta_{p_{l}j}\right]}{\prod_{k=1,k\neq i}^{n}(\Delta_{ij}-\Delta_{kj})}\end{equation}
(for focal pixel $i$, number of points $n$, order $m$) with the operator $\partial_{\Delta}$ defined by Eq. (A29), which can then be used to compute the leakage.
\\For the full-sky discrete estimator, an algebraic solution to the 2-d geometric array is highly involved. Instead, one can compute (by brute force) the numerical power contributions for each derivative term via the Wigner 3\textit{jm}-symbol [16] by treating the $w_{j}$ elements as part of $n$ separate fields, one for each pixel stencil (for the case of two-dimensional square grids):
\begin{equation}\begin{split}(\partial_{n})_{lm}=\sum_{j,l',m'}\frac{4\pi}{N_{\mathrm{pix}}}\sum_{i}w^{(n)}_{j}(\Omega_{i})Y_{l',m'}(\delta\Omega_{j,i})Y_{lm}^{*}(\Omega_{i})\end{split}\end{equation}
where the $\Omega_{j,i}$ refers to the rotations between the focal pixel $i$'s position on the sphere and that of the given neighbouring pixel $j$. It is then possible to calculate the accuracy of the power contribution by measuring the bi-Laplacian (the ``signal'' generated by the derivatives) and the commutator (the ``residual'' generated by the derivatives):
\begin{equation}\mbox{Sig}\leftrightharpoons D^{+}_{\mp2}D^{+}_{0}+D^{-}_{\mp2}D^{-}_{0}=\nabla^{4}\indent\mbox{Res}\leftrightharpoons D^{+}_{\mp2}D^{-}_{0}-D^{-}_{\mp2}D^{+}_{0}\end{equation}
since Eq. (5) can be written in pixel space as
\begin{equation}\begin{split}\nabla^{4}e=\left(\tilde{D}^{+}_{\mp2}D^{+}_{0}+\tilde{D}^{+}_{\mp2}D^{-}_{0}\right)e+\mbox{Err}(e)-\left(\tilde{D}^{+}_{\mp2}D^{-}_{0}-\tilde{D}^{-}_{\mp2}D^{+}_{0}\right)b\\ \nabla^{4}b=\left(\tilde{D}^{+}_{\mp2}D^{+}_{0}+\tilde{D}^{+}_{\mp2}D^{-}_{0}\right)b+\mbox{Err}(b)+\left(\tilde{D}^{+}_{\mp2}D^{-}_{0}-\tilde{D}^{-}_{\mp2}D^{+}_{0}\right)e,\end{split}\end{equation}
where $\mathrm{Err}$ represents remaining error in the calculation from the summation of the ``signal'' and ``residual'' parts. Whilst we can take a simple approximation by assuming leakage into $e$ from $b$ is negligible the full calculation is clearly coupled. A calculation for the residual estimator, using the scheme described in the next subsection, can be seen in Fig. 9. This calculation is then equivalent to defining the pixelization leakage in terms of a convolution operator
\begin{equation}(\partial_{\hat{n}})_{lm}=\sum_{l'm'}\mathcal{W}_{ll'mm'}Y_{l'm'}\end{equation}
where
\begin{equation}\mathcal{W}_{ll'mm'}=\int\left(\sum_{j}w_{j}(\Omega)\right)Y_{l'm'}(\Omega)Y_{lm}(\Omega)d\Omega,\end{equation}
with which one can deconvolve the pixellization leakage. In fact, it is will be useful to derive the more general convolution operator
\begin{equation}\phantom{}_{ss'}\mathcal{W}_{ll'mm'}=\int\partial_{\hat{n}}\left[\phantom{}_{s'}Y_{l'm'}(\Omega)\right]\phantom{}_{s}Y_{lm}(\Omega)d\Omega\end{equation}
if one wishes to perform the exact calculation in the Stokes tensor frame (see Appendix D). Returning from these more general considerations, the residual in the Stokes tensor frame can be approximated by calculating the operators involved in Eq. (3)
\begin{equation}\begin{split}\eth_{-1}\eth_{-2}=\partial_{\theta\theta}-\csc^{2}\theta\partial_{\phi\phi}
+2i\csc\theta\partial_{\theta\phi}+3\cos\theta\csc\theta\partial_{\theta}+2i\cos\theta\csc^{2}\theta\partial_{\phi}-2\hspace{0.4mm}
\\ \bar{\eth}_{1}\bar{\eth_{2}}=\partial_{\theta\theta}-\csc^{2}\theta\partial_{\phi\phi}-2i\csc\theta\partial_{\theta\phi}+3\cos\theta\csc\theta\partial_{\theta}-2i\cos\theta\csc^{2}\theta\partial_{\phi}-2
\end{split}\end{equation}
which are spin-weighted in correspondence with operation on the Stokes tensor instead of the individual $Q$ and $U$ fields, and their discrete analogues, given by
\begin{equation}\tilde{\eth}_{-1}\tilde{\eth}_{-2}f(\Omega)=\eth_{1}\eth_{2}f(\Omega)+R^{+}f(\Omega)\indent\tilde{\bar{\eth}}_{1}\tilde{\bar{\eth}}_{2}f(\Omega)=\bar{\eth}_{1}\bar{\eth}_{2}f(\Omega)+R^{-}f(\Omega),\end{equation}
where the residual operators $R^{\pm}$ are the quantities we will approximate. By taking the lowest order errors in the Taylor series derivation of the first and second order derivatives in 1-d (see Appendix A)
\begin{equation}\begin{split}\frac{f^{j+1}-2f^{j}+f^{j-1}}{\Delta_{x}^{2}}\approx\partial_{xx}f^{j}+\frac{\Delta_{x}^{2}}{12}\partial_{x^{4}}f^{j}\hspace{1.17in}
\\ \frac{f^{j+1}-f^{j-1}}{2\Delta_{x}}\approx\partial_{x}f^{j}+\frac{\Delta_{x}^{2}}{6}\partial_{x^{3}}f^{j}\hspace{1.235in}
\\ \frac{f^{j+1}_{i+1}+f^{j-1}_{i-1}-f^{j+1}_{i-1}-f^{j-1}_{i+1}}{4\Delta_{x}\Delta_{y}}\approx \partial_{xy}f^{j}_{i}+\frac{2}{3}\left(\Delta_{x}^{2}\partial_{x^{3}y}f^{j}_{i}+\Delta_{y}^{2}\partial_{xy^{3}}f^{j}_{i}\right),\end{split}\end{equation}
the operators $R^{\pm}$ are

\begin{equation}\begin{split}R^{\pm}=\Delta_{\theta}^{2}\left(\frac{1}{12}\partial_{\theta^{4}}\pm i\frac{\csc\theta}{3}\partial_{\theta^{3}\phi}+\frac{\cot\theta}{2}\partial_{\theta^{3}}\right)
-\Delta_{\phi}^{2}\left(\frac{\csc^{2}\theta}{12}\partial_{\phi^{4}}\mp\frac{\csc\theta}{3}\partial_{\theta\phi^{3}}\mp2i\cos\theta\csc^{2}\theta\partial_{\phi^{3}}\right)\end{split}\end{equation}
and by substitution into
\begin{equation}\begin{split}\tilde{\nabla}^{4}e=\nabla^{4}e-\sum_{lm}a^{E}_{lm}\left[R^{+}\phantom{}_{-2}Y_{lm}+R^{-}\phantom{}_{2}Y_{lm}\right]+i\sum_{lm}a^{B}_{lm}\left[R^{+}\phantom{}_{-2}Y_{lm}-R^{-}\phantom{}_{2}Y_{lm}\right]
\\ \tilde{\nabla}^{4}b=\nabla^{4}b-\sum_{lm}a^{E}_{lm}\left[R^{+}\phantom{}_{-2}Y_{lm}-R^{-}\phantom{}_{2}Y_{lm}\right]+i\sum_{lm}a^{B}_{lm}\left[R^{+}\phantom{}_{-2}Y_{lm}+R^{-}\phantom{}_{2}Y_{lm}\right]\end{split}\end{equation}
we can calculate the residual approximation. For the case of HEALPix, one has  $\Delta_{\phi}(\theta,\phi)\approx\Delta_{\theta}(\theta,\phi)/\sin\theta$, yielding the contribution at each point in $\theta$:
\begin{equation}\begin{split}\tilde{\nabla}^{4}e=\nabla^{4}e-\sum_{lm}\sqrt{\frac{(l+4)!}{(l-4)!}}\left[a^{E}_{lm}\left(1+\frac{1}{6\sin\theta}\right)Y_{lm}+ia^{B}_{lm}\left(1-\frac{1}{6\sin\theta}\right)Y_{lm}\right]\Delta_{\theta}^{2}
\\ \tilde{\nabla}^{4}b=\nabla^{4}b-\sum_{lm}\sqrt{\frac{(l+4)!}{(l-4)!}}\left[a^{E}_{lm}\left(1-\frac{1}{6\sin\theta}\right)Y_{lm}+ia^{B}_{lm}\left(1+\frac{1}{6\sin\theta}\right)Y_{lm}\right]\Delta_{\theta}^{2}.\end{split}\end{equation}
Alternatively to utilizing Eq. (28), one might simply reconstruct the $Q$ and $U$ fields solely from the sum of the underlying $\nabla^{4}e$-field harmonic coefficients and perform the real-space commutator derivatives. There is also the consistency criterion for calculating $B$-modes on the real-space HEALPix sphere (Fig. 8) which can help inform us as to whether we have calculated a signal which is dominantly numerical noise. Beyond this subsection, $n$ refers to calculation order.

\begin{figure}[H]\centerline{\includegraphics[height=60mm]{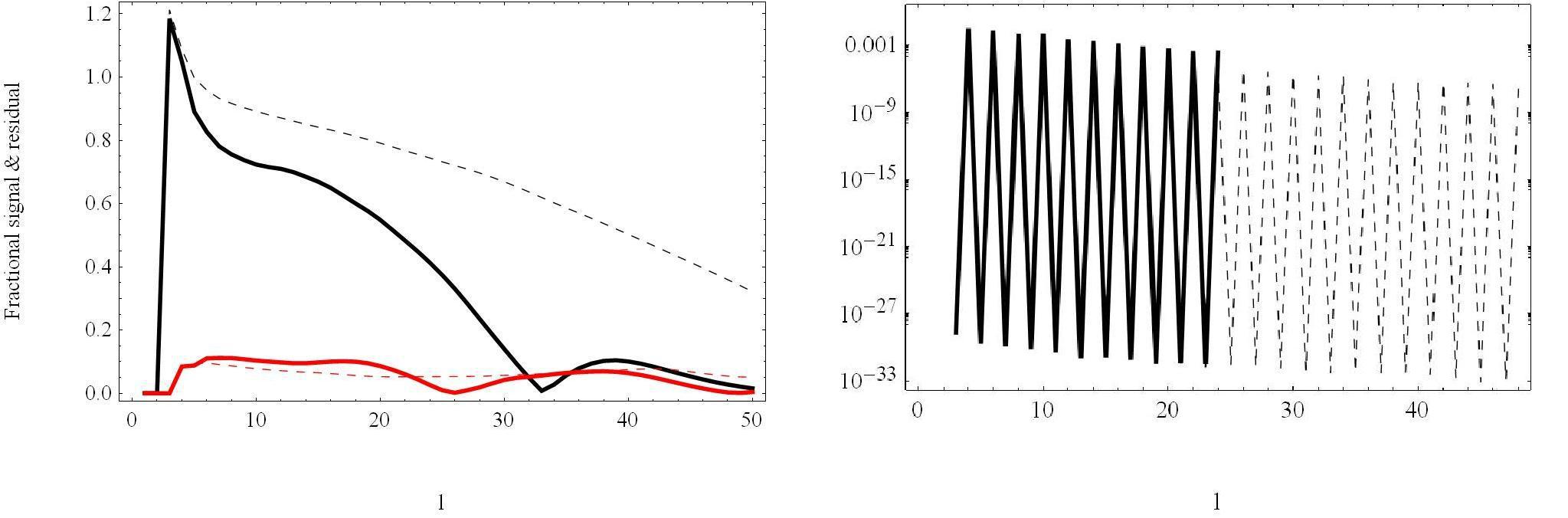}}\caption[]{Left: Fractional signal (black lines) corresponding to $\nabla^4/\tilde{\nabla}^4$ and residual leakage (red lines) of the finite-difference scheme on the HEALPix sphere (as described in sec.III) at order $n=2$; (Thick line, dashed line) corresponds to $N_{\mathrm{side}}=(8, 16)$. There is a characteristic high point at the very low-$l$ scale, followed by an immediate dip in leakage. The first phenomenon is related to the pole problem in our formalism on the HEALPix sphere; the signal starts to peak again at approximately the Nyquist frequency for the map ($\sim 2N_{\mathrm{side}}$). Right: Residuals from individual point $C_l$ maps generated from $a^{E}_{lm}=\delta_{l,10}$, with $N_{\mathrm{side}}=8, 16$ (thick, dashed lines). The leakage has a steep downward gradient across $l$ with the leakage contribution to each multipole oscillating rapidly, the oscillation rate scaling with map resolution.}\end{figure}

\subsection{Software Performance}

\noindent Our analysis so far has been limited to an ideal unmasked sphere; however the real sky is obscured by the galactic plane amongst other foreground sources and will have various contributions of noise to it. We use this subsection to note the performance of the software on a maskless sky against the raw standard pseudo-$C_l$ method implicit in \textit{anafast}, i.e., pure pixellization and finite-difference error, and then show the efficacy of the software in the presence of masking and noise. Of course, whilst testing against standard methods shows that the proposed method works as required, the real test of the software will be against alternative proposals [19, 24, 25, 26, 27, 28, 29, 30] for clean subtraction of the $B$-mode leakage. 

\begin{table}[H]\caption[CMB source maps]{CMB source maps, calculated using the online CMBFAST interface. The source spectra were generated using the online javascript form with default inputs unless otherwise stated, including $K(\eta)_{\mathrm{max}}=3000$, a cosmological constant, Peebles recombination and no 5th dimension. See the online documentation (\url{http://lambda.gsfc.nasa.gov/}) for a discussion of these terms and their implementation.}
\centerline{\begin{tabular}{|c||c||c||c|}
\hline Resolutions & $\Omega_{b}$, $\Omega_{\Lambda}$, $\Omega_{\mathrm{cdm}}$, $\Omega_{\mathrm{hdm}}$, $T_{\mathrm{cmb}}$ (K), & $N_{\nu,\mathrm{massless}}$,  & Tensors? 
\\ $N_{\mathrm{side}}$ generated & $g^{*}_{\mathrm{massive}}$, $H_{0}$ (kms$^{-1}$Mpc$^{-1}$), $Y_{\mathrm{He}}$, $n_{s}$ & $N_{\nu,\mathrm{massive}}$ &
\\ \hline 32, 128 & 0.046, 0.73, 0.224, 0, 2.725, & 3.04, & No 
\\ & 0, 70, 0.24, 0.96 & 0 &
\\ \hline 32, 128 & 0.046, 0.73, 0.224, 0, 2.725, & 3.04, & Yes, 
\\ & 0, 70, 0.24, 0.96 & 0 & $n_{t}=n_{s}-1$
\\ \hline 
\end{tabular}}\end{table}

\noindent Figure 10 compares MasQU calculations on the full unmasked sky to similar HEALPix calculations; on the full sky the harmonic space separation is superior to calculations using the first few stencil sizes; it being the case that with large sky masking the MasQU method is superior (see next subsection), there should exist an intermediate masked area, for a given MasQU calculation order and map resolution, wherein the difference between the HEALPix leakage and MasQU error is minimized. We leave this interesting calculation to a future publication, since for realistic experiments the masking volume will likely exceed such an equilibrium value.

\begin{figure}[h]\centerline{\includegraphics[height=60mm]{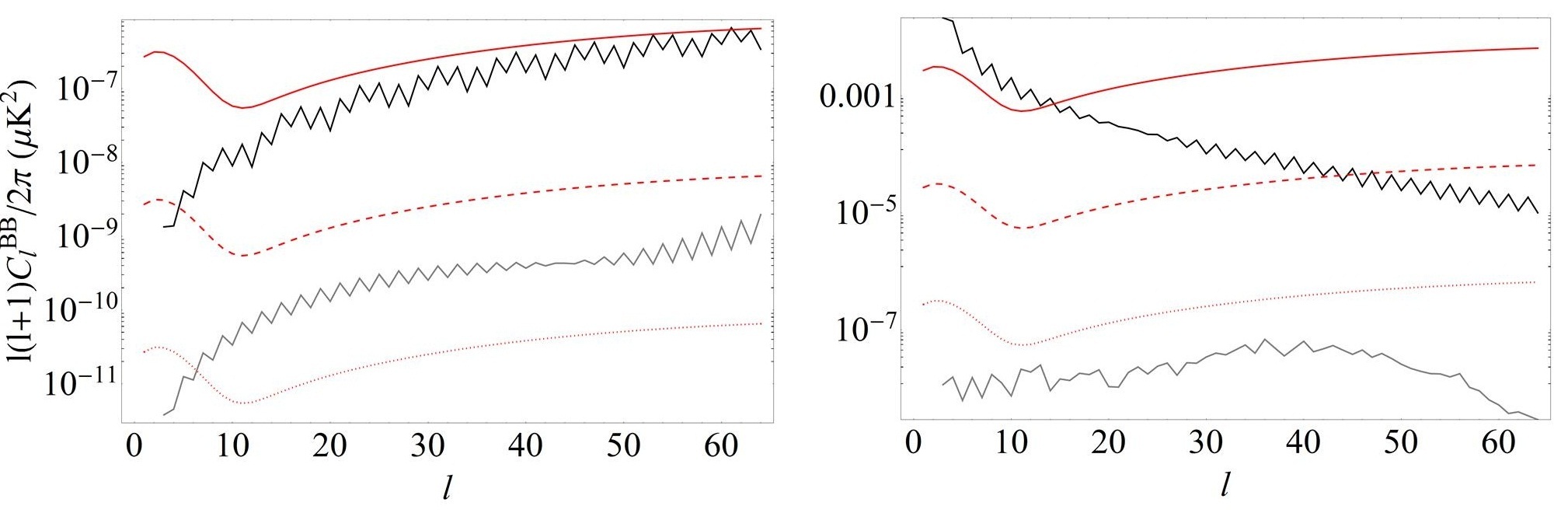}}\caption[]{Maskless calculations, with HEALPix on the left and \begin{math}n=2\end{math} MasQU on the right. For the raw HEALPix calculations (i.e., with no special apodization used), the black line is for 0 iterations while the grey is for 3 iterations of $map2alm$. For the MasQU calculations, the grey line is for a standard calculation, with the pole removed for the black line. The corresponding $\nabla^{4}e$ signal has an rms $\sim2\times10^{-8}$K; the red lines are full-sky $B$-mode spectra from tensor modes corresponding to $r=10^{-5}$, $10^{-7}$, $10^{-9}$ (left-hand diagram) and $r=10^{-1}$, $10^{-3}$, $10^{-5}$ (right-hand diagram).}\end{figure}

\noindent Pole removal (Fig. 11) has the effect of lowering power across scales, with the decrement most noticeable at low $l$. This was implemented by removing the ($n+1$) rings immediately outside each pole. For completeness, we also show the difference between operating in the full-sky and flat-sky formalisms; it should not be surprising that the errors in the flat-sky approximation are at their largest at low $l$.

\begin{figure}[h]\centerline{\includegraphics[height=70mm,width=190mm]{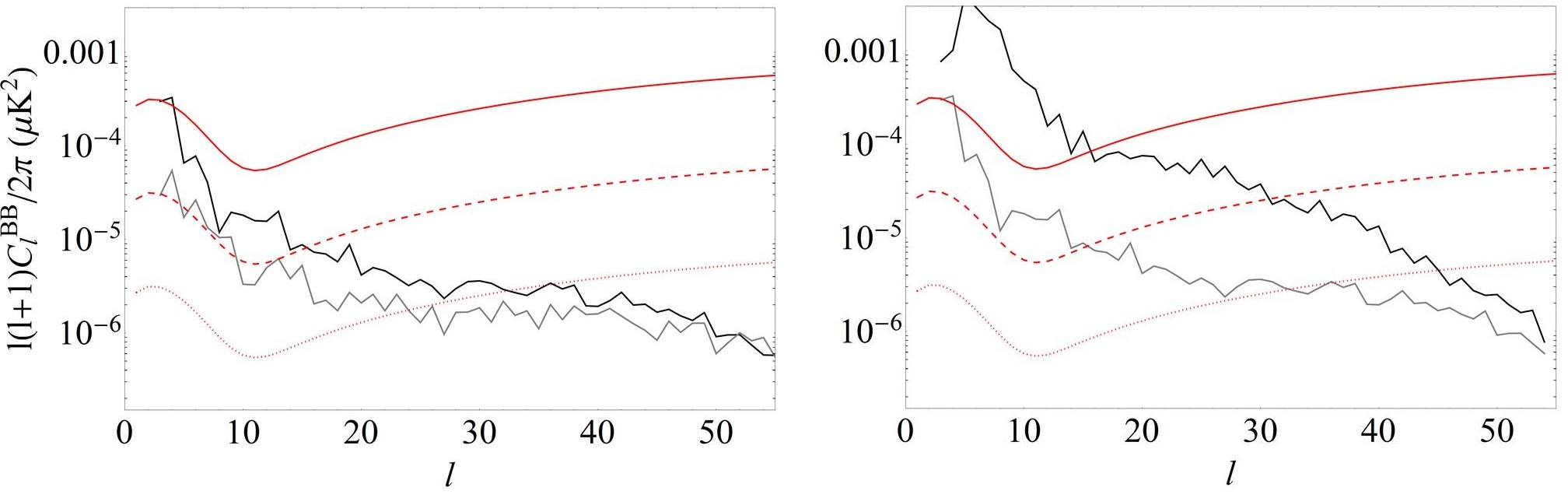}}\caption[]{Left diagram: $n=2$ and $n=4$ (black, grey) maskless $B$-modes from the $N_{\mathrm{side}}=32$, $r=0$ map, with the pole removed. Right: \begin{math}n=2\end{math} calculations with the flat-sky and full-sky operators (black, grey). In both plots, the red lines are $B$-mode spectra from tensor modes corresponding to $r=10^{-2}$, $10^{-3}$, $10^{-4}$.}\end{figure}

\subsection{Masking}

\begin{figure}[h]\centerline{\includegraphics[height=100mm]{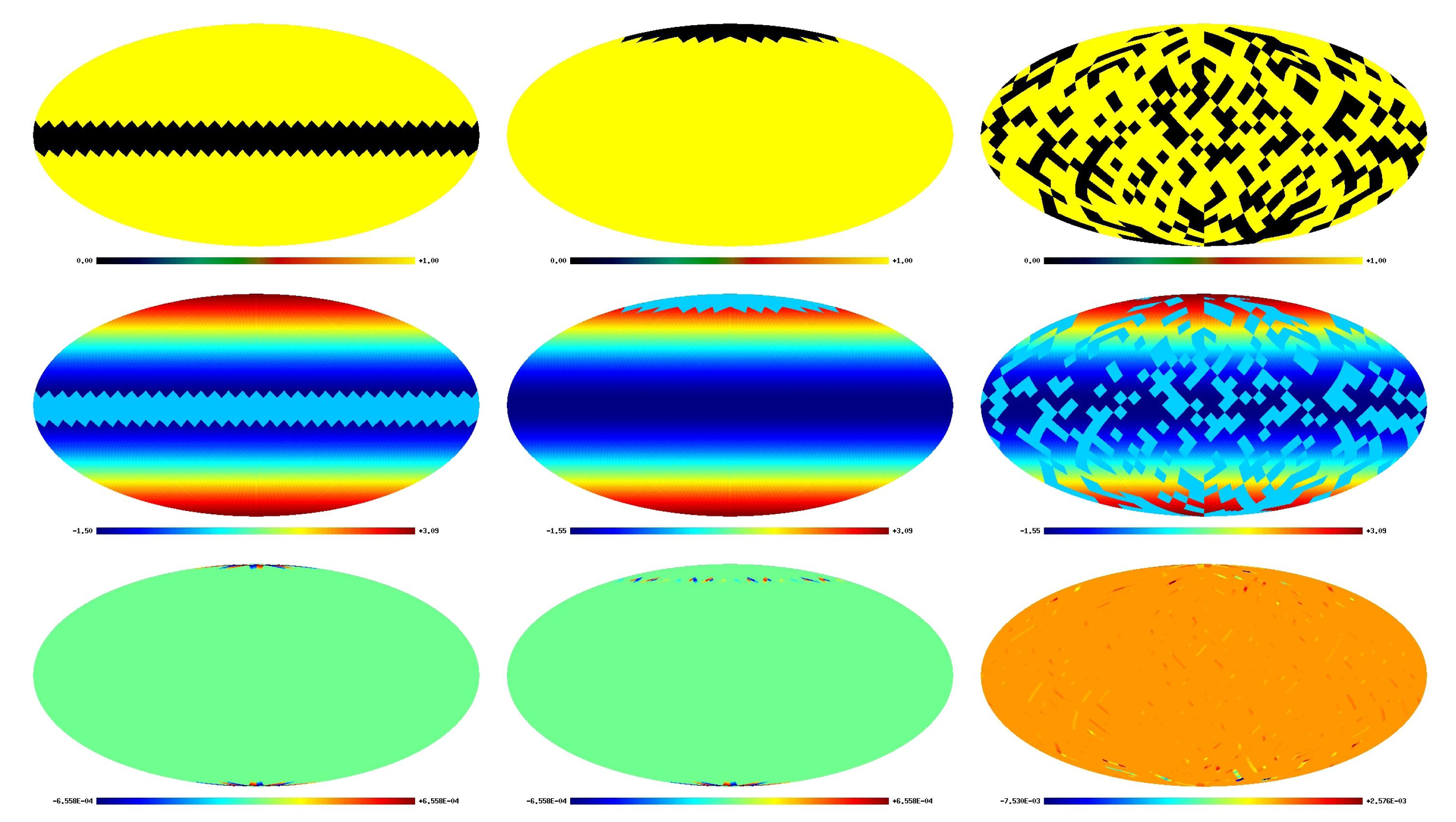}}\caption[]{
Rows, top-to-bottom: Masking schemes, $\nabla^{4}e$ maps and $\nabla^{4}b$ maps, at $n=6$ and $N_{\mathrm{side}}=32$ for $a^{E}_{lm}=\delta_{l2}\delta_{m0}$, $a^{B}_{lm}=0$. Left to right: the equatorial mask ($f_{\mathrm{sky}}\approx0.83$), polar mask ($f_{\mathrm{sky}}\approx0.96$) and random mask ($f_{\mathrm{sky}}\approx0.64$). The source function was chosen for the smallness of the pole problem errors, but nonetheless the masking errors are smaller or of magnitude that contributed by the pole problem; specifically, the absolute maximum values in the $\nabla^{4}b$ maps are, from left to right, ($\sim6.6\times 10^{-4},\sim6.6\times10^{-4},\sim7.5\times10^{-3}$).}\end{figure}

\begin{figure}[h]\centerline{\includegraphics[height=120mm]{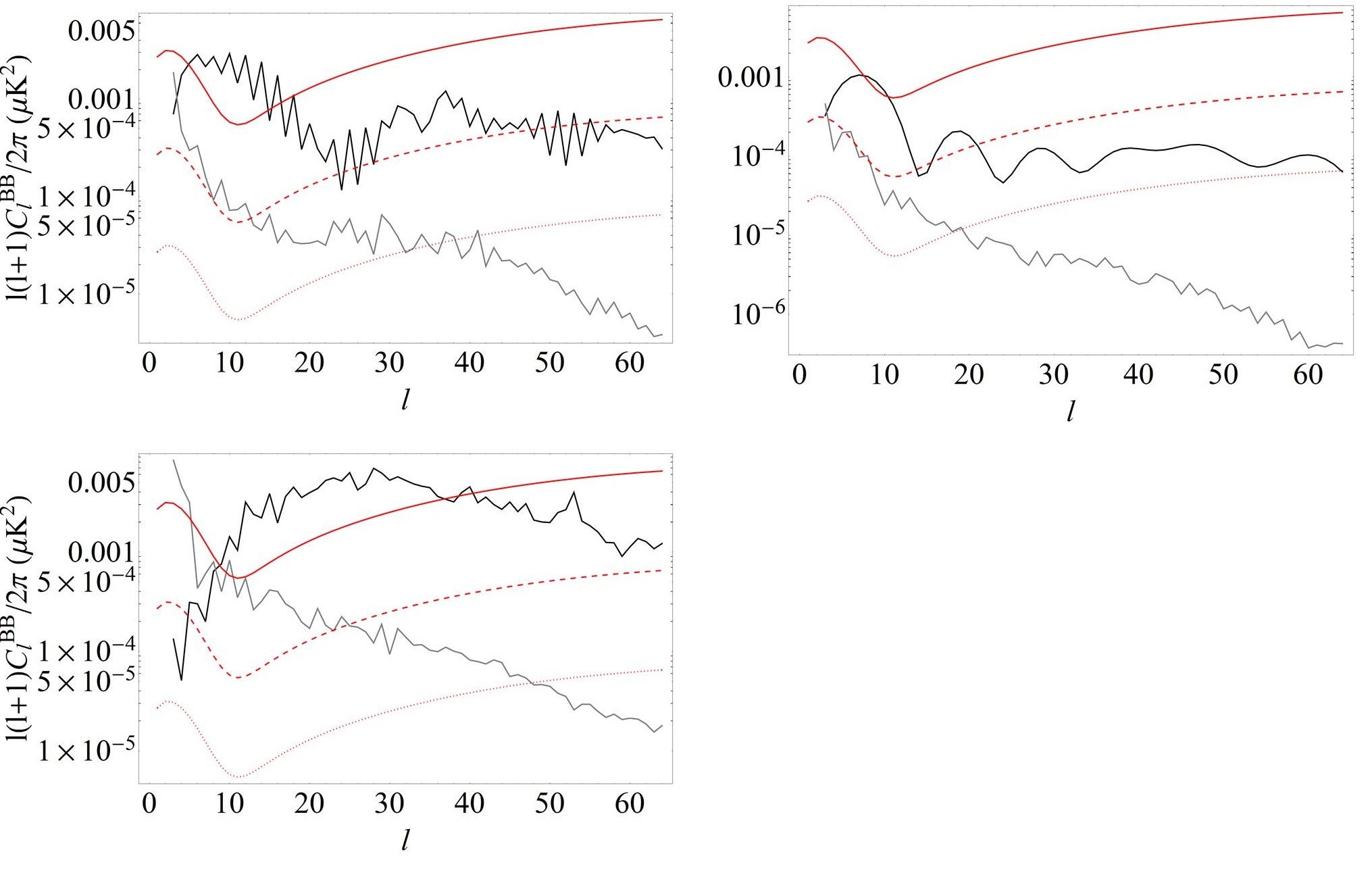}}\caption[]{$B$-mode \begin{math}C_{l}\end{math}s from our fiducial $B$-mode-free maps as reconstructed by the 3-iterations HEALPix method (black) or an \begin{math}n=2\end{math} MasQU calculation (grey), for equatorial, polar and random masks. In all the plots, the red lines are $B$-mode spectra from tensor modes corresponding to $r=10^{-1}$, $10^{-2}$, $10^{-3}$.}\end{figure}

\noindent In the case of masking, analysis was performed using 3 basic types of mask (Fig. 12): equatorial, polar and random masks. In the masked case differences are found as would be expected; the pole problem particularly affects the power at around $l=2$  by boosting it significantly. Focussing on the no-tensors maps, at multipoles higher than $l\sim10$ MasQU performs considerably better than the raw pseudo-$C_l$ calculations for masked HEALPix schemes, by about 1 to 3 orders of magnitude; in most models gravitational lensing rather than primordial modes dominate the foreground polarization from $l\sim150$, meaning we have a large $l$-range where MasQU is advantageous for calculating $B$-modes. The smoothness of the masked calculations is in contrast to that of underlying functional discontinuities. The effect of a shelf discontinuity itself is shown in Fig. 14. As expected (since the underlying approximation to the signal is an interpolating polynomial), the software actually performs worse with a larger stencil when in the presence of a discontinuity. For the CMB, discontinuities will mostly be contibuted by point sources on the sky; these will need to be masked away using a source catalogue. By extension, the software could be used as a tool to search for discontinuities, as discussed in [31].

\begin{figure}[h]\centerline{\includegraphics[height=30mm]{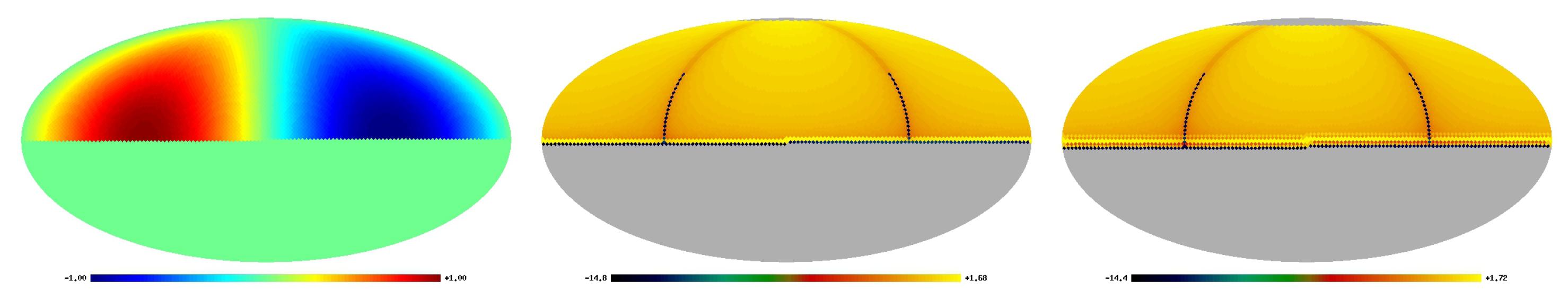}}\caption[]{(Logarithmic) code performance in the presence of shelf discontinuities; the $B$-modes are calculated from a $Q=\sin\theta\sin\phi$, $U=0$ pair of $N_{\mathrm{side}}=128$ maps with a cut-off to $Q=0$ in the southern hemisphere at the equator. Without a $U$ signal, one should have no $B$-modes. Left-to-right: Original $Q$ map, $n=2$ and $n=4 \nabla^{4}b$ maps. The map value ranges for each image are $(-1,1)$, $(-14.8,1.68)$ and $(-14.4,1.72)$ respectively. Notice that there is a blow-up in errors for calculations across the discontinuity, as expected; the error actually gets worse and effects a larger region for larger stencils, and also scales with the magnitude of the discontinuity. This general behaviour also follows for point discontinuities.}\end{figure}

\subsection{Noise Performance}

\noindent A number of simple noise models (white Gaussian, anisotropic uncorrelated and pixel-to-pixel correlated) are also analysed. The Gaussian model was calculated using the HEALPix random number-generating subroutine \textit{planck-rng} in harmonic space rather than pixel space, since one can then look at the effect of the dominant noise scale on the calculations. For the crude non-Gaussian-white-noise maps, the anisotopic noise map was constructed from a pixel-level Gaussian noise map with a small functional direction-dependence imposed on it whilst the pixel-correlated signal was constructed from a pixel-level Gaussian noise map with reflection symmetry imposed between the north and south hemispheres. Figure 15 presents power spectra for the realistic $B$-mode-less CMB maps with Gaussian noise added, where variations have been made in the mean value of the noise (at 10\%, 1\% and 0.1\% the mean signal values of the noiseless CMBFAST-generated $Q$ and $U$ maps, i.e., $\left<X_{\mathrm{noise}}\right>\propto\left<X_{\mathrm{sig}}\right>$ where $X\in\{Q,U\}$) 
and in the scaling of the noise (via a cut-off in the number of multipoles generated for the noise maps). At high magnitude and large $l$,
since the noise translates to a collection of point discontinuities in real space, the sum of the derivatives of the summed signal and noise maps can be expected to feature more point source errors. The addition of the Gaussian noise models serves to boost $B$-mode power fairly consistently across \textit{l} up to the $l_{\mathrm{max}}$ value that sets the smallest scale for noise; a lower-\textit{l} noise mode cut-off $l_{\mathrm{max}}$ results in a drop in signal power boosting for $l>l_{\mathrm{max}}$, but still with a significant contribution. The pixel-correlated power boost is almost indistinguishable from the pure Gaussian boost; in contrast, the anisotropic boost is dependent on the scaling of the direction-dependent noise. The MasQU method can then be seen to be somewhat sensitive to noise; one would want a good understanding of the systematic and foreground noise properties in order to effectively purify the $B$-modes.

\begin{figure}[h]\centerline{\includegraphics[height=70mm,width=190mm]{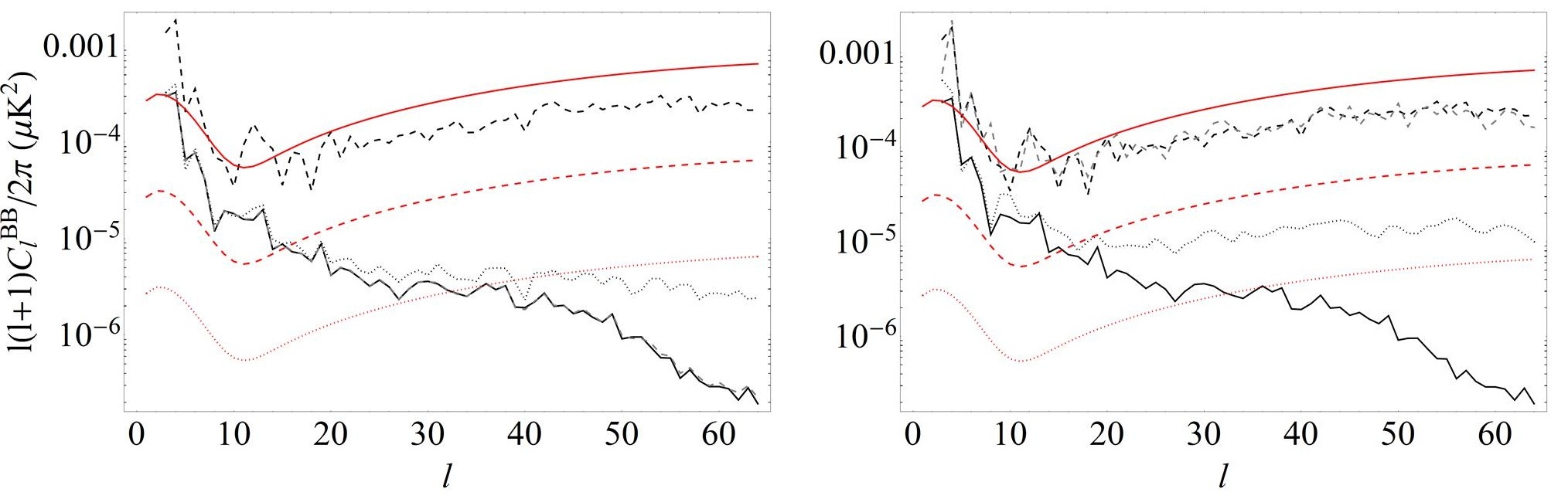}}\caption[]{Left: \begin{math}n=2\end{math} maskless $B$-modes from the $r=0$ map with resolution $N_{\mathrm{side}}=32$, with noise added to the underlying $Q$, $U$ maps and the pole removed. The Gaussian noise models in $Q$ and $U$ are calculated from a Gaussian distribution harmonic coefficients and normalized such that it's mean is some proportion of the signal mean, i.e., $\left<N\right>=\alpha\left<S\right>$. Left plot: The thick black line is our noiseless model, whilst the scaling for the dashed black, dotted black, and dashed grey lines are $\alpha=0.1$, $0.01$, $0.001$ respectively. Right plot: The thick black line is the noiseless model, with the dashed black, dotted black, and dashed grey lines for Gaussian, anisotropic, and pixel-correlated noise models respectively, all with $\alpha=0.1$. In both plots, the red lines are $B$-mode spectra from tensor modes corresponding to $r=10^{-2}$, $10^{-3}$, $10^{-4}$. }\end{figure}

\subsection{Apodization}

\noindent Any masking scheme will ensure that the \textit{l}-space power of an underlying signal is redistributed across multipoles, since there are some scales which will contribute less depending on the mask. Gibbs overshooting will also be a problem; boundary effects ensure that ``ringing'' occurs in any harmonic approximation. This is usually dealt with by apodizing the signal --- applying a tapering weight to each pixel contribution to the power, such that the apodization function tends smoothly to zero at the boundary.
\\In the $E$- and $B$-mode case, an optimal apodization scheme in the pseudo-\begin{math}C_{l}\end{math} formalism has been derived by Smith \& Zaldarriaga [19]. It can be shown by operating $\eth$ on a window function representing the masking that in the analytic case passing to the scalar field is equivalent to the ``pseudo-$C_l$-with-counterterms'' method. Specifically,
\begin{equation}b_{lm}=\frac{i}{2}\int d\Omega (Q+iU)(W\phantom{}_{2}Y_{lm}^{*}+\frac{2\phantom{}_{1}W^{*}\phantom{}_{1}Y_{lm}^{*}}{\sqrt{(l-1)(l+2)}}+\frac{\phantom{}_{2}W^{*}Y_{lm}^{*}}{\sqrt{(l-1)l(l+1)(l+2)}}+\mbox{c.c.})\end{equation}
and equivalently for the $e$-term, where $\phantom{}_{s}W=\eth^{s}W$. The optimal apodization weights for a minimum variance estimator are calculated by minimizing the difference between the band-limited pseudo-$C_l$ estimator
\begin{equation}\tilde{C}_{\mathrm{band}}=\sum_{ij}d_{i}W_{i}C_{ij}^{\mathrm{band}}W_{j}d_{j},\end{equation}
constructed using the mask map $W_{i}$ where the band-limiting is defined by a binning of multipoles $l$, and the optimal estimator
\begin{equation}\tilde{O}_{\mathrm{band}}=\sum_{ij}d_{i}(C^{-1}C^{\mathrm{band}}C^{-1})_{ij}d_{j}\end{equation}
where $d$ is a data vector and $C$ is a covariance matrix (which may be band-limited); this is 
then calculated by conjugate gradient inversion (varying the spin-weighted $W$ components independently) and is only equivalent to the polarization modes in the mean of multipoles \textit{m} (with subsequent loss of phase information). The method used in [19] is somewhere between ``pure'' (without mode-mixing) and optimal. By contrast, our real fields $\nabla^{4}e$ and $\nabla^{4}b$ are by construction pure. An application of this method to our scheme is left to a future paper.

\subsection{Leakage from Realistic Surveys}

\noindent In order to get a rough idea of the leakage improvement the software can bring to real data, the leakage for sky coverages in a simplified model of the \textit{E and B Experiment} (EBEX [32]) survey and the bounds that may be set on the tensor-to-scalar ratio $r$ in these simplistic cases is calculated; the algorithm is also performed on a mask of similar area centred on the equator. In the past the signal-to-noise ratio $S/N$ has been too low to perform differencing on the real sky, but such projects are designed to improve $S/N$, making such calculations plausible. The fiducial model is the same as the parameters set for the previous no-tensors analysis (Table 1).
\\The EBEX survey is a balloon-borne polarimeter for probing the sky with a resolution of less than 8 arcminutes at frequency bands centered at 150, 250, 350, and 450 GHz. The sky patch covered by the $\sim$1300-detector instrument corresponds to 350 square degrees. The EBEX region is an approximately square patch, corresponding to Fig. 9 in Stivoli et al [33]. In particular, MasQU calculations are performed that are equal roughly to the EBEX ``sky''and ``ground'' coverages found in Stivoli et al, for the fiducial $B$-mode-free model adding uncorrelated Gaussian noise, at the levels 3.2 $\mu$K and 0.9 $\mu$K for the sky and ground coverages respectively\footnote{From correspondences with the EBEX team.}, in both $Q$ and $U$ (i.e., $\left<X\right>_{\mathrm{noise,\phantom{ }``sky''}}=3.2\mu$K and $\left<X\right>_{\mathrm{noise,\phantom{ }``ground''}}=0.9\mu$K where $X=\{Q,U\}$). This is performed on an $N_{\mathrm{side}}=128$ resolution HEALPix map --- less than the resolution capable by EBEX, but more than enough to capture the essential low-$l$ polarization information relevant to tensor modes. The procedure starts by smoothing the signal + noise map with a Gaussian kernel, and calculate the $E$- and $B$-mode spectra from the resulting smoothed map (Fig. 16); the noise model is approximated by performing the same derivatives on a smoothed Gaussian field. Since the convolution of the map with a Gaussian function smooths high-resolution variation, this amounts to a low-frequency pass filter. On the HEALPix sphere, the decreasing resolution per ring will serve to increase low-$l$ power in the calculations. Since one is not using the full sky it is advantageous to rotate the survey region to the equator, which would ameliorate such a problem. This is because the pixel distributions in the polar cap and equatorial regions differ significantly.
\\Mock likelihoods for the tensor-to-scalar ratio $r$ from Gaussian priors (centred on $r=0$ with $(\mathrm{min}, \mathrm{max}) = (0,0.33)$) are also computed, using CosmoMC [34] and $\sim 200,000$ likelihood space samplings: all cosmological parameters are held constant\footnote{Baryon density $\Omega_{b}h^{2}=0.0223$, dark matter density $\Omega_{dm}h^{2}=0.105$, optical depth $\tau=0.09$, curvature $\Omega_{k}=0$ and dark energy equation-of-state parameter $w=-1$.} except for varying $r$, the scalar and tensor spectral indices $n_{s}$, $n_{t}$ and the superhorizon power of the scalar perturbations $\log A_{s}$, centred on 0.95, 0 and 3 respectively. This of course assumes that quantities such as the reionization optical depth are known perfectly; for a realistic analysis one would have to perform the MCMC calculations over a higher-dimensional
space that includes such parameters as variables. Thus if the noise model is well-known, then the MasQU method provides an excellent improvement in the mock surveys over standard harmonic methods. Specifically, since real-space derivatives obey linearity, one can in principle approximate the noiseless map by removing the $E$- and $B$-modes calculated by a full real-space noise model. Given that the noise models from projects such as WMAP are computed initially as sky maps, this would have the advantage of separating out point sources before smoothing of the map is performed.

\begin{figure}[H]\centerline{\includegraphics[height=100mm]{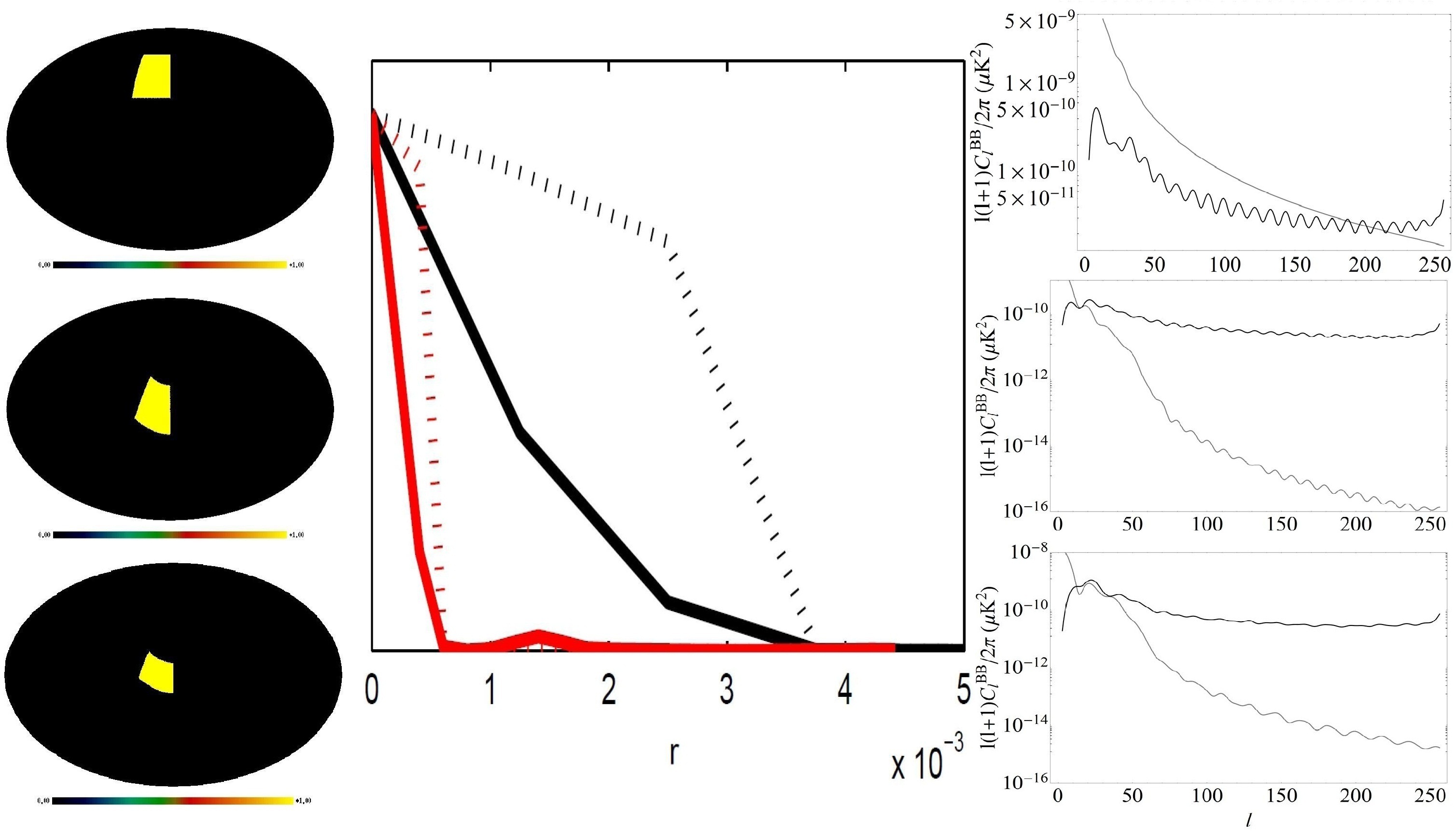}}\caption[]{Left column, top-to-bottom: Mock EBEX ground survey, rotated ground survey, and the rotated sky survey masks. The ground and sky surveys are imposed with a 0.9 $\mu\mathrm{K}$ and 3.2 $\mu\mathrm{K}$ smoothed Gaussian noise component, representing detector error. Right column, $B$-mode spectra corresponding to the survey regions in the left column; the black line is a HEALPix calculation, the grey an $n=2$ MasQU calculation. These results compare favourably with the $B$-mode residual in more detailed EBEX analysis [35], although the noise model used in the analysis here is far more simplistic. Central column: Maximum likelihood calculations of $r$ for the middle row mask according to the text; the black lines are the HEALPix calculations, with the MasQU calculations in red (solid lines are the fully marginalized posteriors, dotted lines the relative mean likelihoods).}\end{figure}

\section{Conclusions}

\noindent We have presented a software package for performing derivatives on the HEALPix grid, in the presence an arbitrary mask. The setup given is in fact quite malleable and can be adapted to any pathology-free grid, for any collection of derivatives. This is particularly useful in the case of CMB polarization studies, where we can use derivative operators to separate out the $E$ and $B$ polarization modes and avoid mode-mixing as is problematic in standard harmonic methods.
\\In the particular case of the operators used for CMB analysis with HEALPix, we found that polarization mode numerical operations at the pole were highly sensitive to error in the component derivatives, leading to very large numerical noise at the lower order differencing stencils. This was found to be ameliorable in any of four ways (i) throwing away pixels at the pole --- while this removes information, there is an automatic reduction in contamination; (ii) applying larger differencing stencils around the pole --- here we lose no information but have a slow convergence rate for large multipole sources; (iii) rotating the grid in $\theta$, and then rotating back the calculated field values --- this suffers from ringing which can be reduced by essentially doubling the entire spherical calculation, or by using a tapering function; or (iv) ``rotated'' sampling --- whilst we lose no information, and in theory have a quicker convergence at large multipoles, this requires very accurate $\mathrm{map}\rightarrow a_{lm}\rightarrow\mathrm{map}$ reconstructions in order to be useful, something that standard HEALPix methods do not supply.
\\In some senses, na\"{i}ve speed comparisons for the MasQU software against HEALPix harmonic methods are somewhat inappropriate; whilst HEALPix performs considerably faster per standard run, MasQU requires calculation of weights only once per masking scheme,map resolution and stencil size --- any further calculations using the same three criteria requires only a trivial summation at each pixel of the recorded weights, regardless of the underlying sky map.
\\Work on the software is ongoing. Further studies will involve quantifying statistics of MasQU maps, the correct apodization of masked maps, and a full exposition of the discontinuity-finding aspects of the formalism. A beta version of MasQU is available at \url{https://sourceforge.net/projects/masqu/}. Use of the MasQU package for producing results should be acknowledged in any forthcoming publication. Bugs or technical issues should be emailed to the first author.

\begin{acknowledgements}

\noindent The authors would like to thank Marc Kamionkowski, Radek Stompor and the members of the EBEX team for useful discussions. JB was supported by an STFC studentship, and AHJ and DN acknowledge support of STFC grants.

\end{acknowledgements}


\noindent $^{*}$Electronic address: \ttfamily j.bowyer07@imperial.ac.uk
\\ \rmfamily $^{\dag}$Electronic address: \ttfamily a.jaffe@imperial.ac.uk
\\ \rmfamily $^{\ddag}$Electronic address: \ttfamily d.novikov@imperial.ac.uk
\\ \rmfamily$[$1$]$ Guth, A. [1981], Phys. Rev. D \textbf{23} 347
\\ $[$2$]$ Seljak, U. \& Zaldarriaga, M. [1997], Phys. Rev. Lett \textbf{78} 2054
\\ $[$3$]$ Tegmark, M. \& de Oliveira Costa, A. [2001], Phys. Rev. D, \textbf{64}, 063001
\\ $[$4$]$ Jaffe, A., Kamionkowski, M., \& Wang, L. [2000], Phys.Rev. D \textbf{61}, 083501
\\ $[$5$]$ Ames, W. [1977], \textit{Numerical Methods for Partial Differential Equations}, Academic Press
\\ $[$6$]$ Gorski, K. \textit{et al} [2005], Astrophys. J. \textbf{622} 759
\\ $[$7$]$ Bunn, E., arXiv:0811.0111
\\ $[$8$]$ Doroshkevich, A.G. \textit{et al} [2005], Int. J. Mod. Phys. D \textbf{14} 275
\\ $[$9$]$ The Planck Collaboration, ArXiv:astro/0604069
\\ $[$10$]$ Kovac, J.M. \textit{et al} [2002], Nature \textbf{420}, 772
\\ $[$11$]$ Hinshaw, G. \textit{et al}, \textit{Nine-Year Wilkinson Microwave Anisotropy Probe (WMAP) Observations: Cosmological Parameter Results}, arXiv:1212.5226
\\ $[$12$]$ Kamionkowski, M., Koswosky, A. \& Stebbins, A. [1997], Phys. Rev. D \textbf{55} 7368
\\ $[$13$]$ Goldberg, J. \textit{et al} [1967], J.Math.Phys. \textbf{8} 2155
\\ $[$14$]$ Li, J. [2005], J. Comp \& Ap. Math., \textbf{183}, 29
\\ $[$15$]$ Marcus, M. \& Minc, H. [1992], \textit{A Survey of Matrix Theory and Matrix Inequalities}, Dover
\\ $[$16$]$ Varshalovich, D.A., Moskalev, A.N. \& Khersonkii, V.K. [1988], \textit{Quantum Theory of Angular Momentum}, World Scientific
\\ $[$17$]$ Press \textit{et al} [2007], \textit{Numerical Recipes, Third Ed.}, Cambridge University Press
\\ $[$18$]$ Anderson, E. \textit{et al} [1999], \textit{LAPACK User's Guide, Third Ed.}, SIAM
\\ $[$19$]$ Smith, K. \& Zaldarriaga, M. [2007], Phys. Rev. D \textbf{76} 043001
\\ $[$20$]$ Seljak, U. \& Zaldarriaga, M [1996], Astrophys. J. \textbf{469} 437
\\ $[$21$]$ LAMBDA (\textit{Legacy Archive for Microwave Background Data Analysis}) webpage, \url{http://lambda.gsfc.nasa.gov/}
\\ $[$22$]$ Gorski, K. \textit{et al}, arXiv:astro-ph/9905275
\\ $[$23$]$ Arfken, G. [1985], \textit{Mathematical Methods for Physicists}, Academic Press
\\ $[$24$]$ Efstathiou, G. [2006], Mon. Not. Roy. Astron. Soc. \textbf{370} 343
\\ $[$25$]$ Zhao, W. \& Baskaran, D. [2010], Phys. Rev. D \textbf{82}, 023001
\\ $[$26$]$ Bunn, E., arXiv:1008.0827
\\ $[$27$]$ Kim, J. \& Naselsky, P. [2010], A\&A \textbf{519} A104
\\ $[$28$]$ Kim, J., arXiv:1010.2636
\\ $[$29$]$ Cao, L. \& Fang, L. [2009], Astrophys. J. \textbf{706} 1545
\\ $[$30$]$ Geller, D. \textit{et al} [2008], 	Phys. Rev. D \textbf{78} 123533
\\ $[$31$]$ Bowyer, J. \& Jaffe, A.H., arXiv:1011.1791
\\ $[$32$]$ Oxley, P. \textit{et al} [2004], Proc. SPIEInt. Soc. Opt. Eng. \textbf{5543} 320
\\ $[$33$]$ Reichborn-Kjennerud \textit{et al} [2010], Conference proceedings for SPIE Millimeter, Submillimeter, and Far-Infrared Detectors and Instrumentation for Astronomy V
\\ $[$34$]$ Lewis, A. \& Bridle, S. [2002], Phys. Rev. D \textbf{66} 103511
\\ $[$35$]$ Stivoli \textit{et al} [2010], Mon. Not. Roy. Astron. Soc. \textbf{408} 2319
\\ $[$36$]$ Milne-Thompson, L. M., [1933] \textit{The Calculus of Finite Differences}, MacMillan and Company
\\ $[$37$]$ Horn, R. \& Johnson, C. [1985], \textit{Matrix Analysis}, Cambridge University Press
\\ $[$38$]$ Smith, K. [2006], Phys. Rev. D \textbf{74} 083002
\\ $[$39$]$ Gradshteyn, I.S., \& Ryzhik, I.M. [2007], \textit{Table of Integrals, Series and Products, 7th Ed.}, Academic Press

\appendix
\section{A General Finite-Difference Scheme}

\noindent This Appendix briefly expands on the finite-difference scheme; most of the results here are taken from [5, 14, 36, 37].
\\A derivative (of order $m$) of a function at a given pixel $i$ can be computed as the sum of weighted values of the function at a surrounding sample of pixels $j$:
\begin{equation}\partial_{x^{m}}f_{i}\approx\sum_{j}^{\mathrm{pixels}}w^{(m)}_{ij}f_{j}\end{equation}
where \textit{w} is the weight matrix. For example, the canonical examples of a finite-difference scheme with uniform spacing can be derived starting with a one-dimensional Taylor expansion on an infinite regular grid with separation $\Delta$
\begin{equation}\begin{split}f_{i\pm1}=f_{i}\pm\Delta\partial_{x}f_{i}+\Delta^{2}\frac{\partial_{xx}f_{i}}{2!}\pm\cdots=\sum^{\infty}_{r=0}\frac{(\pm\Delta)^{r}\partial_{x^{r}}f_{i}}{r!}\end{split}\end{equation}
to yield for the 1st and 2nd derivatives
\begin{equation}\partial_{x}f_{i}\approx\frac{f_{i+1}-f_{i-1}}{2\Delta}+\mathcal{O}(\Delta^{2}),\indent\partial_{xx}f_{i}\approx\frac{f_{i+1}+f_{i-1}-2f_{i}}{\Delta^{2}}+\mathcal{O}(\Delta^{2})\end{equation}
corresponding to weight matrices
\begin{equation}w^{(1)}_{ij}=\left(\begin{array}{ccc}-\frac{1}{2\Delta}&0&\frac{1}{2\Delta}\\ -\frac{1}{2\Delta}&0&\frac{1}{2\Delta}
\\ \vdots &\vdots &\vdots\end{array}\right),\indent w^{(2)}_{ij}=\left(\begin{array}{ccc}\frac{1}{\Delta^{2}}&-\frac{2}{\Delta^{2}}&\frac{1}{\Delta^{2}}\\ \frac{1}{\Delta^{2}}&-\frac{2}{\Delta^{2}}&\frac{1}{\Delta^{2}}
\\ \vdots &\vdots &\vdots\end{array}\right).\end{equation}
Since the focal pixel in this case is the central pixel $f_{i}$ between $f_{i\pm1}$, this is usually referred to as the ``second-order central difference scheme''. If the grid is finite, the first and last rows in the weight matrix must correspond to ``forward'' and ``backward'' difference schemes respectively, where the focal pixel is $f_{i\mp1}$ in the second-order case. Such schemes can be constructed using the same Taylor analysis as in the central difference case.
\\The following more general formalism shall be constructed to approximate the derivatives at a single pixel; in that sense we shall utilise the weight vector $w_{j}$ (at fixed but unlabelled pixel $i$) instead of $w_{ij}$. For the set of all pixels on a pixellated grid, the weight vector $w_{j}$ corresponds to rows of $w_{ij}$.
\\A general finite-difference method, for any number of regular or irregular pixel schemes can be derived using interpolating polynomials. In the Lagrange basis, a 1-d polynomial interpolating a set of $n$ datapoints can be written as
\begin{equation}f(x)\approx\sum_{i=1}^{n}f_{i}L_{i}(x),\end{equation}
where the $f_{i}$ are the datavalues at each point $x_{i}$ and the Lagrange basis polynomial is
\begin{equation}L_{i}(x)=\prod_{1\leq j\leq n; j\neq n}\frac{x-x_{j}}{x_{i}-x_{j}}\end{equation}
such that on a sample of $n$ pixels we have
\begin{equation}\left(\begin{array}{ccc}L_{1}(x_{1}) & \cdots & L_{n}(x_{1}) \\ \vdots &  & \vdots \\ L_{n}(x_{1}) & \cdots & L_{n}(x_{n})\end{array} \right)\left(\begin{array}{c} f_{1} \\  \vdots \\ f_{n}\end{array} \right)=\left(\begin{array}{c} f_{1}\\ \vdots \\ f_{n}\end{array} \right)\end{equation}
where clearly $L_{i}(x_{j})=\delta_{ij}$. It can be shown that for a given nondegenerate distribution of points, the Lagrange basis polynomial both exists and is unique. Interpolation problems over \textit{n} pixels arranged on a 1-d grid can often be expressed using a geometric progression matrix, also called the ``Vandermonde'' matrix [15]:
\begin{equation}v=\left(\begin{array}{ccc}x_{1}^{0} & \cdots & x_{1}^{n-1} \\ \vdots &  & \vdots \\ x_{n}^{0} & \cdots & x_{n}^{n-1}\end{array} \right)\end{equation}
which is constructed by rewriting the interpolating polynomial in the monomial basis, by a simple rearrangement of terms in Eq. (A6):
\begin{equation}L_{i}=c_{1}+c_{2}x_{i}+\cdots=\sum_{r=1}c_{r}x_{i}^{r-1}\end{equation}
and then filling the rows of the matrix $v$ as appropriate, such that
\begin{equation}\left(\begin{array}{ccc}x_{1}^{0} & \cdots & x_{1}^{n-1} \\ \vdots &  & \vdots \\ x_{n}^{0} & \cdots & x_{n}^{n-1}\end{array} \right)\left(\begin{array}{c} c_{1} \\  \vdots \\ c_{n}\end{array} \right)=\left(\begin{array}{c} L_{1}\\ \vdots \\ L_{n}\end{array} \right).\end{equation}
Using this, a scheme for approximating the derivatives of the interpolating polynomial can be constructed; if we write Eq. (A10) in the compact form $vc=L$, noting that the positive definiteness of $v$ implies that $(v^{-1})^{T}=(v^{T})^{-1}$, we can define a unique unspecified array $\alpha$ which solves the transpose equation
\begin{equation}v^{T}\alpha=L,\end{equation}
corresponding to the summation
\begin{equation}L_{i}=\sum_{r=1}\alpha_{r}x_{r}^{i}.\end{equation}
The linear equation for these interpolation weights is then
\begin{equation}\left(\begin{array}{ccc}x_{1}^{0} & \cdots & x_{n}^{0} \\ \vdots &  & \vdots \\ x_{1}^{n-1} & \cdots & x_{n}^{n-1}\end{array} \right)\left(\begin{array}{c} \alpha_{1} \\  \vdots \\ \alpha_{n}\end{array} \right)=\left(\begin{array}{c} L_{1}\\ \vdots \\ L_{n}\end{array} \right).\end{equation}
By isolating a single pixel of interest, with position $x$, and replacing the positions of the surrounding pixels $x_{i}$ with the position difference
\begin{equation}\Delta_{i}=x_{i}-x\end{equation}
one is led to an equation (with an unspecified array $W$)
\begin{equation}v'=WL'\end{equation}
defined in powers of $\Delta_{i}$
\begin{equation}\left(\begin{array}{ccc}\Delta_{1}^{0} & \cdots & \Delta_{n}^{0} \\ \vdots &  & \vdots \\ \Delta_{1}^{n-1} & \cdots & \Delta_{n}^{n-1}\end{array} \right)\left(\begin{array}{c} W_{1} \\ \vdots \\ W_{n}\end{array} \right)=\left(\begin{array}{c} 0!L_{1} \\ \vdots \\ (n-1)!L^{(n-1)}_{1} \end{array}\right),\end{equation}
which can be modified for calculating the differencing weights (where $L_{i}^{(n)}$  is a shorthand for the $n^{\mathrm{th}}$ derivative of $L_{i}$). We refer to the array $v'$ as the ``differenced Vandermonde'' array. Let us now consider a more general Taylor series than Eq. (A2) for our polynomial function $L$ in order to modify Eq. (A16) for calculating differencing weights:
\begin{equation}L_{i\pm n}=\sum^{\infty}_{r=0}\frac{(\pm\Delta_{n})^{r}L_{i}^{(r)}}{r!}.\end{equation}
In a matrix format, since our summation is realistically limited to the $(n-1)^{\mathrm{th}}$ derivative this is none other than
\begin{equation}\left(\begin{array}{ccc}\frac{(\pm\Delta_{1})^{0}}{0!} & \cdots & \frac{(\pm\Delta_{1})^{n-1}}{(n-1)!} \\ \vdots &  & \vdots \\ \frac{(\pm\Delta_{n})^{0}}{0!} & \cdots & \frac{(\pm\Delta_{n})^{n-1}}{(n-1)!}\end{array} \right)\left(\begin{array}{c} L_{i}^{(0)} \\ \vdots \\ L_{i}^{(n-1)}\end{array} \right)=\left(\begin{array}{c} L_{i\pm1} \\ \vdots \\ L_{i\pm n} \end{array}\right).\end{equation}
By inverting the matrix Eq. (A18), we can calculate each of the derivatives of the polynomial $L$. It is then clear that in order to isolate a particular derivative $m$, one must append the left-hand-side of the inverted equation with a Kronecker delta thus:
\begin{equation}\left(\begin{array}{c} 0!L_{i}^{(0)}\delta_{0,m} \\ \vdots \\ (n-1)!L_{i}^{(n-1)}\delta_{n-1,m}\end{array} \right)=\left(\begin{array}{ccc}(\pm\Delta_{1})^{0} & \cdots & (\pm\Delta_{1})^{n-1} \\ \vdots &  & \vdots \\ (\pm\Delta_{n})^{0} & \cdots & (\pm\Delta_{n})^{n-1}\end{array} \right)^{-1}\left(\begin{array}{c} L_{i\pm1} \\ \vdots \\Ll_{i\pm n} \end{array} \right)\end{equation}
where we have made use of the array $(\pm v'^{T})^{-1}$. Correspondingly, we modify Eq. (A16) for the same purpose:
\begin{equation}\left(\begin{array}{ccc}\Delta_{1}^{0} & \cdots & \Delta_{n}^{0} \\ \vdots &  & \vdots \\ \Delta_{1}^{n-1} & \cdots & \Delta_{n}^{n-1}\end{array} \right)\left(\begin{array}{c} w_{1}^{(m)} \\ \vdots \\ w_{n}^{(m)}\end{array} \right)=\left(\begin{array}{c} L_{1}\delta_{m,0} \\ \vdots \\ (n-1)!L^{(n-1)}_{1}\delta_{m,n-1} \end{array} \right)\end{equation}
in order to separate out the derivative of interest. By extension, replacing the Kronecker term via $\delta_{m,r}\rightarrow\delta_{m||m',r}$ which evaluates to $1$ if $either$ $m$ or $m'$ is equal to $r$, will calculate the weights required to compute the summed derivative $L^{(m)}+L^{(m')}$. Thus the array $W$ is related to the vector weights $w^{(m)}$ by
\begin{equation}W_{i}=\sum_{j=0}^{n-1}w_{i}^{(j)}.\end{equation} 
One can then solve for the weight vector by isolating the derivative polynomial of choice. For example, the central difference system for a regular grid, where $\Delta=1$, can be obtained from
  \begin{equation}\left(\begin{array}{ccc}1 & 1 & 1 \\ -1 & 0 & 1 \\ 1 & 0 & 1\end{array} \right)\left(\begin{array}{c} w^{(1)}_{0} \\ w^{(1)}_{1} \\ w^{(1)}_{2}\end{array} \right)=\left(\begin{array}{c} 0 \\ 1 \\ 0\end{array} \right),\end{equation}
or a backward difference equation for the second derivative via
   \begin{equation}\left(\begin{array}{ccc}1 & 1 & 1 \\ 0 & 1 & 2 \\ 0 & 1 & 4\end{array} \right)\left(\begin{array}{c} w^{(2)}_{0} \\ w^{(2)}_{1} \\ w^{(2)}_{2}\end{array} \right)=\left(\begin{array}{c} 0 \\ 0 \\ 2\end{array} \right),\end{equation}
with results in agreement with the known results. A range of standard regular finite-difference weights is displayed in Table III.

\begin{table}[here] \caption[]{2-, 3-, and 4-point equidistant 1st-order difference equations.}
\begin{tabular}{|c||c||c||c||c||c|}\hline Points  & 2-pt-bwd & 1-pt bwd & Central & 1-pt-fwd & 2-pt-fwd \\ 
\hline 2  & - & \begin{math}\frac{F_{i}-F_{i-1}}{\Delta}\end{math} & - & \begin{math}\frac{F_{i+1}-F_{i}}{\Delta}\end{math} & -  \\
\hline 3  & \begin{math}\frac{-3F_{i-2}+4F_{i-1}-F_{i}}{2\Delta}\end{math} & - & \begin{math}\frac{-F_{i-1}+F_{i+1}}{2\delta}\end{math} & - & \begin{math}\frac{F_{i}-4F_{i+1}+3f_{i+2}}{2\Delta}\end{math} \\
\hline 4 & \begin{math}\frac{-2F_{i-2}-3F_{i-1}+6F_{i}-F_{i+1}}{6\Delta}\end{math} & - & - & - & \begin{math}\frac{F_{i-1}-6F_{i}+3F_{i+1}+2F_{i+2}}{6\Delta}\end{math} \\
\hline
\end{tabular}\end{table}

\noindent The extension of this to an irregular grid merely requires the reparameterization
\begin{equation}\Delta_{ij}=x_{j}-x_{i}.\end{equation}
So for the derivatives at a pixel \textit{i} one solves
\begin{equation}\left(\begin{array}{ccc}\Delta_{i,1}^{0} & \cdots & \Delta_{i,n}^{0} \\ \vdots &  & \vdots \\ \Delta_{i,1}^{n-1} & \cdots & \Delta_{i,n}^{n-1}\end{array} \right)\left(\begin{array}{c}w_{i,1}^{(m)} \\ \vdots \\ w_{i,n}^{(m)}\end{array}\right)=\left(\begin{array}{c}0!\delta_{m,0} \\ \vdots \\ (n-1)!\delta_{m,n-1}\end{array}\right).\end{equation}
A general solution to this set-up for a derivative of order $m$ can be determined by constructing the $LU$ decomposition of the inverse of the trace of the Vandermonde matrix into lower- and upper-triangular arrays $\Lambda$ and $\Upsilon$ respectively:
\begin{equation}(v^{T})^{-1}=\Lambda\Upsilon\end{equation}
where
\begin{equation}\begin{split}\Lambda=\left(\begin{array}{ccccc}1 & -x_{1}& x_{1}x_{2} & -x_{1}x_{2}x_{3} & \hdots 
\\0&1&-(x_{1}+x_{2})& x_{1}x_{2}+x_{2}x_{3}+x_{3}x_{1}&\hdots \\ 0&0&1&-(x_{1}+x_{2}+x_{3})&\hdots \\ \vdots &\vdots &\vdots &\vdots &\ddots
\end{array}\right)\hspace{0.3cm}
 \\ \Upsilon=\left(\begin{array}{cccc}1&0&0&\hdots
\\ \frac{1}{x_{1}-x_{2}}&\frac{1}{x_{2}-x_{1}}&0&\hdots
\\ \frac{1}{(x_{1}-x_{2})(x_{1}-x_{3})}&\frac{1}{(x_{2}-x_{1})(x_{2}-x_{3})}&\frac{1}{(x_{3}-x_{1})(x_{3}-x_{2})}&\hdots
\\ \vdots &\vdots &\vdots &\ddots
\end{array}\right).
\end{split}\end{equation}
The analogous decomposition for the differenced Vandermonde matrix leads to the general solution for an $n$-point finite-difference scheme in 1 dimension
\begin{equation}w_{ij}^{(m)}=\frac{(\partial_{\Delta})^{m}\left[\prod_{l=1,p_{l}\neq i}^{n}\Delta_{p_{1}j}\cdots\Delta_{p_{l}j}\right]}{\prod_{k=1,k\neq i}^{n}(\Delta_{ij}-\Delta_{kj})}\end{equation}
where $i$ is the focal pixel and $j$ denotes weights applied to the stencil pixels and we have defined the following operator:
\begin{equation}\partial_{\Delta}=\sum_{j=1}^{n}\partial_{\Delta_{ij}}\end{equation}
yielding for example
\begin{equation}\begin{split}\partial_{xx,4-pt}f^{i}=\frac{2(\Delta_{i 2}+\Delta_{i 3}+\Delta_{i 4})}{(\Delta_{i 1}-\Delta_{i2})(\Delta_{i 1}-\Delta_{i 3})(\Delta_{i 1}-\Delta_{i 4})}f^{1}+
\frac{2(\Delta_{i 1}+\Delta_{i 2}+\Delta_{i 4})}{(\Delta_{i 2}-\Delta_{i 1})(\Delta_{i 2}-\Delta_{i 3})(\Delta_{i 2}-\Delta_{i 4})}f^{2}\\+\frac{2(\Delta_{i 1}+\Delta_{i 3}+\Delta_{i 4})}{(\Delta_{i 3}-\Delta_{i 1})(\Delta_{i 3}-\Delta_{i 2})(\Delta_{i 3}-\Delta_{i 4})}f^{3}+\frac{2(\Delta_{i 1}+\Delta_{i 2}+\Delta_{i 3})}{(\Delta_{i 4}-\Delta_{i 1})(\Delta_{i 4}-\Delta_{i 2})(\Delta_{i 4}-\Delta_{i 3})}f^{4}.\end{split}\end{equation}
If the determinant of an $n$-point Vandermonde array is labelled $\mbox{Det}[v]_{n}$, then by utilizing standard linear algebra techniques it can be shown that
\begin{equation}\mbox{Det}[v]_{n}=\prod_{j=2}^{n}(x_{j}-x_{1})\mbox{Det}[v]_{n-1}.\end{equation}
By iteration, the determinant of the Vandermonde array is then
\begin{equation}\mbox{Det}[v]_{n}=\prod_{1\leq i<j\leq n}^{n}(x_{j}-x_{i})\end{equation}
which, for a given unknown $x_{j}$, with positions $x_{i}$ known, is precisely the factorised interpolating polynomial. It is can then be seen that the Lagrange polynomials can be identified with the determinants for Vandermonde arrays determined by the pixel sample surrounding pixel $j$ (where we have dropped the $n$):
\begin{equation}L_{i}(x_{j})=\left.\frac{\mbox{Det}[v_{i}]}{\mbox{Det}[v]}\right|_{j}\end{equation}
where the subscripts on $v$ indicate that the Vandermonde matrix defined at pixel $j$ has the column $i$ replaced by a column of undetermined values for $x$, i.e., for the one-dimensional case,
\begin{equation}v_{2}=\left(\begin{array}{ccccc}x_{1}^{0}&x^{0}&x_{3}^{0}& \cdots & x_{n}^{0} \\ \vdots & \vdots & \vdots & \ddots &\vdots 
\\ \vdots & \vdots & \vdots & \ddots & \vdots
\\ x_{1}^{n-1}&x^{n-1} & x_{3}^{n-1} & \cdots & x_{n}^{n-1}\end{array}\right),\end{equation}
since the Lagrange polynomial is merely the weighted sum of the unique polynomials at each point on the grid. Thus in the $d$-dimensional case, one would also have
\begin{equation}f_{j}\approx\sum^{n}_{i}f_{i}\left.\frac{\mbox{Det}[v^{(d)}_{i}]}{\mbox{Det}[v^{(d)}]}\right|_{j}\end{equation}
where we have to specify the correct form of the $d$-dimensional geometric array. One corrollary of the interpolating polynomial being in the form of a geometric array is that we can immediately test for the existence of a unique polynomial (in the powers we have specified in the geometric array) with roots at each pixel. Such a polynomial does not exist if the geometric array is singular. Similarly, one can construct general solutions for the derivative weights in $d$ dimensions as functions of the determinant of $v'$:
\begin{equation}w^{(m^{a})}_{ij}=\frac{(\partial_{\Delta^{a}})^{m_{a}}P_{ij}^{(d)}}{\mbox{Det}[v'^{(d)}]}\end{equation}
where the derivative operator is
\begin{equation}(\partial_{\Delta^{a}})^{m_{a}}=\overbrace{\partial_{\Delta_{e_{1}}}}^{\times m_{1}}\cdots\overbrace{\partial_{\Delta_{e_{d}}}}^{\times m_{d}}\end{equation}
with $\Delta^{a}=\left(\Delta_{e_{1}}, \Delta_{e_{2}}, \cdots, \Delta_{e_{d}}\right)$ the vector of position differences in each of the dimensions $e_{i}$, $m^{a}$ the vector enumerating the derivative orders in each dimension and $P^{(d)}$ is a multinomial of $\Delta$ which we can construct from the $(n-1)^{\mathrm{th}}$ minors of our $n\times n$ geometric array. The determinant depends on the geometry of the geometric array.
\\By looking at the residual $R$ on the Taylor series truncation
\begin{equation}L_{i\pm n}=\sum^{n}_{r=0}\frac{(\pm\Delta_{n})^{r}L_{i}^{(r)}}{r!}+R_{n+1},\indent |R_{n+1}|\approx\frac{\Delta_{n}^{n+1}L_{i}^{(n+1)}}{(n+1)!}\end{equation}
which can be generalized to higher dimensions by utilizing the Hessian form of the Taylor expansion (A2)
\begin{equation}L_{i\pm1}=L_{i}\pm\Delta^{a}\partial_{a}L_{i}+\frac{\Delta^{a}\Delta^{b}\partial_{ab}L_{i}}{N_{ab}}+\cdots,\end{equation}
where the summation convention is assumed and $\partial_{ab\cdots}$ are the partial derivative tensors of Hessian type i.e., in 2 dimensions
\begin{equation}\partial_{a}=\left(\begin{array}{c}\partial_{x}\\ \partial_{y}\end{array}\right),\indent \partial_{ab}=\left(\begin{array}{cc}\partial_{xx}&\partial_{xy}\\ \partial_{yx}&\partial_{yy}\end{array}\right)\end{equation}
and $N_{ab\cdots}$ is a numerical factor corresponding to the factorial terms in Eq. (A2), it can be shown that the error on our general $n$-point $d$-dimensional finite-difference scheme for a square array is of order
\begin{equation}\mbox{Err}\sim\mathcal{O}\left(\left[\prod_{i\neq j}\Delta_{ij}\right]^{1/d}\right),\end{equation}
where we recover $\mathcal{O}(\Delta^{2})$ for 1-point radius regular central schemes in $d$-dimensions, or $\mathcal{O}(\Delta^{2n})$ for $n$-point radius regular central schemes. For non-pathological functions such as polynomials, the accuracy of the differencing scheme generally improves with the number of points used (Fig. 1).
\\Finally, we should note that there may yet be some value in using geometric arrays featuring inverse powers. Whilst the Taylor series does not permit negative powers in the expansion, the Laurent expansion [23] of a complex function
\begin{equation}f(x)=\sum_{n=-\infty}^{\infty}a_{n}(x-b)^{n}\end{equation}
(where the $a_n$ terms are constants and $b$ is a point on the complex plane) does, suggesting that one can deal with a pixel at a coordinate singularity by applying the hyperbolic part of the interpolation at that pixel. This would, for example, correspond to the following algorithm for the differencing weights to approximate the 1st-order derivative on a 1-d grid:
  \begin{equation}\left(\begin{array}{ccc} \Delta_{1}^{-1} & \Delta_{2}^{-1} & \Delta_{3}^{-1} \\1 & 1 & 1 \\ \Delta_{1} & \Delta_{2} & \Delta_{3}\end{array} \right)\left(\begin{array}{c} w^{(1)}_{0} \\ w^{(1)}_{1} \\ w^{(1)}_{2}\end{array} \right)=\left(\begin{array}{c} 0 \\ 0 \\ 1\end{array}\right).\end{equation}

\section{Differencing at the Pole}

\noindent It is initially surprising that one has to be careful when performing differencing at the pole, since the pole is an artefact of the coordinate scheme rather than any intrinsic geometrical problem. Appendix B of [38] suggests a covariant finite-differencing scheme to deal with this issue, which will be implemented in a future version of the software; for now, the methods already in place are both simple and effective enough.
\\When performing a differencing calculation it is important to choose a single consistent coordinate basis, to avoid computational difficulties. For the standard calculations across the remainder of the sphere, the derivative calculations are simplest in spherical polar coordinates ($\theta$, $\phi$); alternatives, such as utilizing a two-dimensional Cartesian coordinate system or performing the calculation in the embedding space $\mathbb{R}^3$ are more involved (the former for example, requires two coordinate grids and a coordinate patch at the equator).
\\In the spherical polar basis, coordinate singularities at the poles $\theta=0$ and $\theta=\pi$ present problems in constructing numerically-stable differencing stencils about either pole. While HEALPix never does sample the pole directly, the ``nearest-neighbour'' pixel stencils about the pole are severely deformed from the standard square-array setup in these coordinates (see Fig. 17). Constructing linear Vandermonde equations corresponding to the various stencil geometries at each order $n$ in the polar region is not only a algorithmic nuisance, but the deformed geometries are also much less numerically stable than the standard approximately-square stencils. Alternatively, since the vector separation between any two points can be calculated with ease one might choose utilize the two-point difference equations. However, this is not sufficient for the accurate calculation of second-order differential operators.
\\At and around the poles, it may be worth performing more involved calculations for the sake of accuracy. In this subsection of the Appendix, methods of dealing with the pole are briefly discussed.

\begin{figure}[H]\centerline{\includegraphics[scale=0.9]{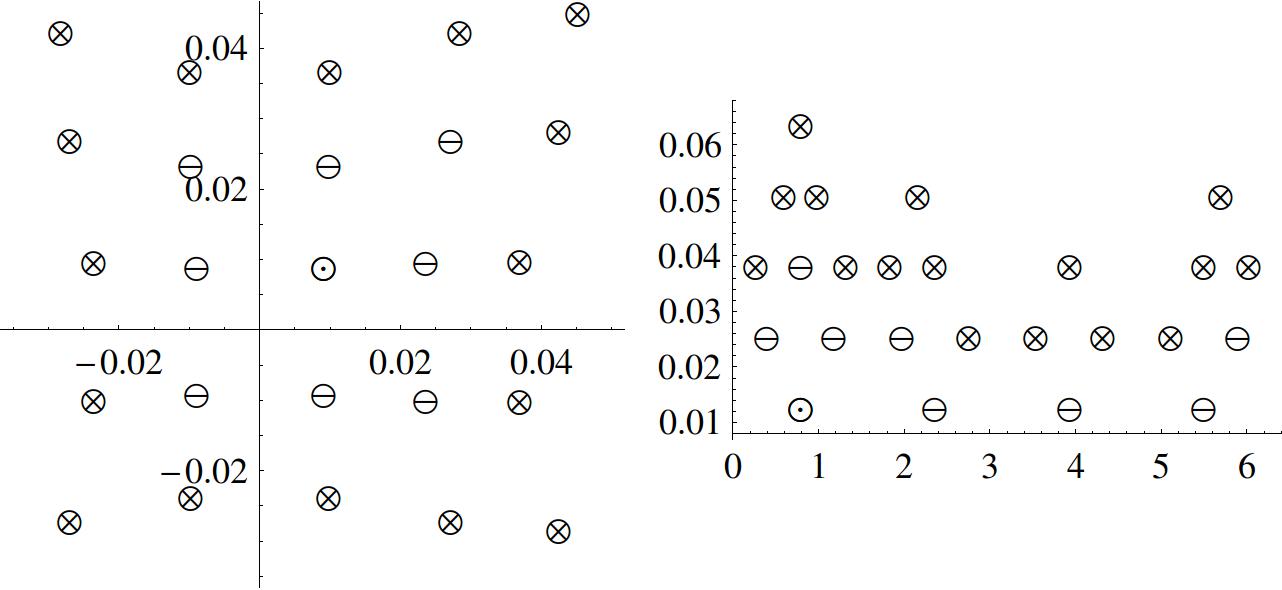}}\caption[Stencil]{Nearest-neighbour pixels distributions at the HEALPix pole in Cartesian ($x$, $y$) and spherical polar ($\theta$, $\phi$) coordinate bases (left and right images, respectively). The circles represent pixel positions for a pole-covering $n=4$ stencil; the subset of circles with a horizontal bar are the neighbouring-pixel positions for an $n=2$ stencil, with the dotted circle the central pixel position, and the cross circles are the remaining $n=4$ pixel positions. In the spherical coordinate system these form highly irregular stencils. Note that the $\Delta\phi$ resolution problem does not exist for the Cartesian system.}\end{figure}

\subsection{Pole-covering schemes}

\noindent Since central-differencing schemes are the most accurate kind of differencing schemes and the easiest to set up in HEALPix, it is preferable to be able to use stencils that cover the pole. One approach might be to transform coordinates, and perform differencing in the new coordinate system. Transforms which place the pole at infinity are ruled out, since an initially central-differencing scheme becomes an outer-difference scheme in the new coordinate basis (the central pixel is the one nearest the pole in spherical coordinates).
\\When constructing new pole-crossing coordinates \begin{math}(\lambda_{1},\lambda_{2})\end{math}, it is desirable that $(\lambda_{1},\lambda_{2})$ are real-valued and unique for each point on $\mathbb{S}^{2}$, and that the parabolicity in \begin{math}\theta\end{math} and the double-valuedness in $\phi$ (at \begin{math}\phi=0/2\pi\end{math}) are not manifested in the new coordinate system.
This essentially limits the form the coordinate transforms can take to variants on the standard Cartesian transformation
\begin{equation}\begin{split}x=\cos\phi\sin\theta,\indent y=\sin\phi\sin\theta;
\indent \theta=\sin^{-1}(\sqrt{x^{2}+y^{2}}),\indent\phi=\tan^{-1}(y/x).\end{split}\end{equation}

\noindent An alternative might be to numerically compute a Jacobian 
\begin{equation}J=\left(\begin{array}{cc}\partial_{e_{1}}\theta & \partial_{e_{2}}\theta \\ \partial_{e_{1}}\phi & \partial_{e_{2}}\phi \end{array}\right)\end{equation}
to fit to a regular grid --- taking a distribution of pixels and calculating a coordinate system numerically that transforms those pixels into a regular distribution $(e_{1},e_{2})$; one then doesn't need to know what the coordinate transform \textit{is}. The remainder of the method is then to calculate the derivatives of $Q$ with respect to $(e_{1},e_{2})$ by using the canonical regular grid weights, and convert between partial derivatives of $(e_{1},e_{2})$ and \begin{math}(\theta,\phi)\end{math} using the calculated Jacobian. 
\\The problems with using coordinate transforms are twofold: first, analytic transforms repeat some calculations (such as determining the stencil geometry), whilst numerical transforms (such as the Jacobian technique, with which one might map to a perfectly regular grid) are not computationally cheap; second, testing becomes more involved --- if one uses a map constructed from polynomials on the sphere, such as the sum of its harmonic coefficients, then the functional behaviour of the transformed map close to the pole is often difficult for a polynomial to model using an interpolant (for example, for a polynomial $f(e_1,e_2)$ on a pathological coordinate system, a transformation to a pathology-free coordinate system $e_3,e_4$ invariably leaves the transformed functional $f(e_3,e_4)$ description with equivalent functional pathologies). Granted one does not expect real data, with all its nuances, to be in polynomial form in any particular coordinate system. In any case, a simpler and more wieldy alternative is found in rotating the sphere.

\subsection{Rotating the sphere}

\noindent As mentioned, rotation provides a good option for dealing with polar issues. Rotating in $\phi$ allows one to ameliorate the ``$\Delta\phi$ problem'' and double the stencil size, whilst rotating from the pole to the equator not only removes all the polar problems listed in the main text, but allows one to simply re-use weights already calculated at the equator. An example of the MasQU result by rotating the sphere in $\theta$ is seen in Fig. 19.

\begin{figure}[H]\centerline{\includegraphics[height=80mm,width=135mm]{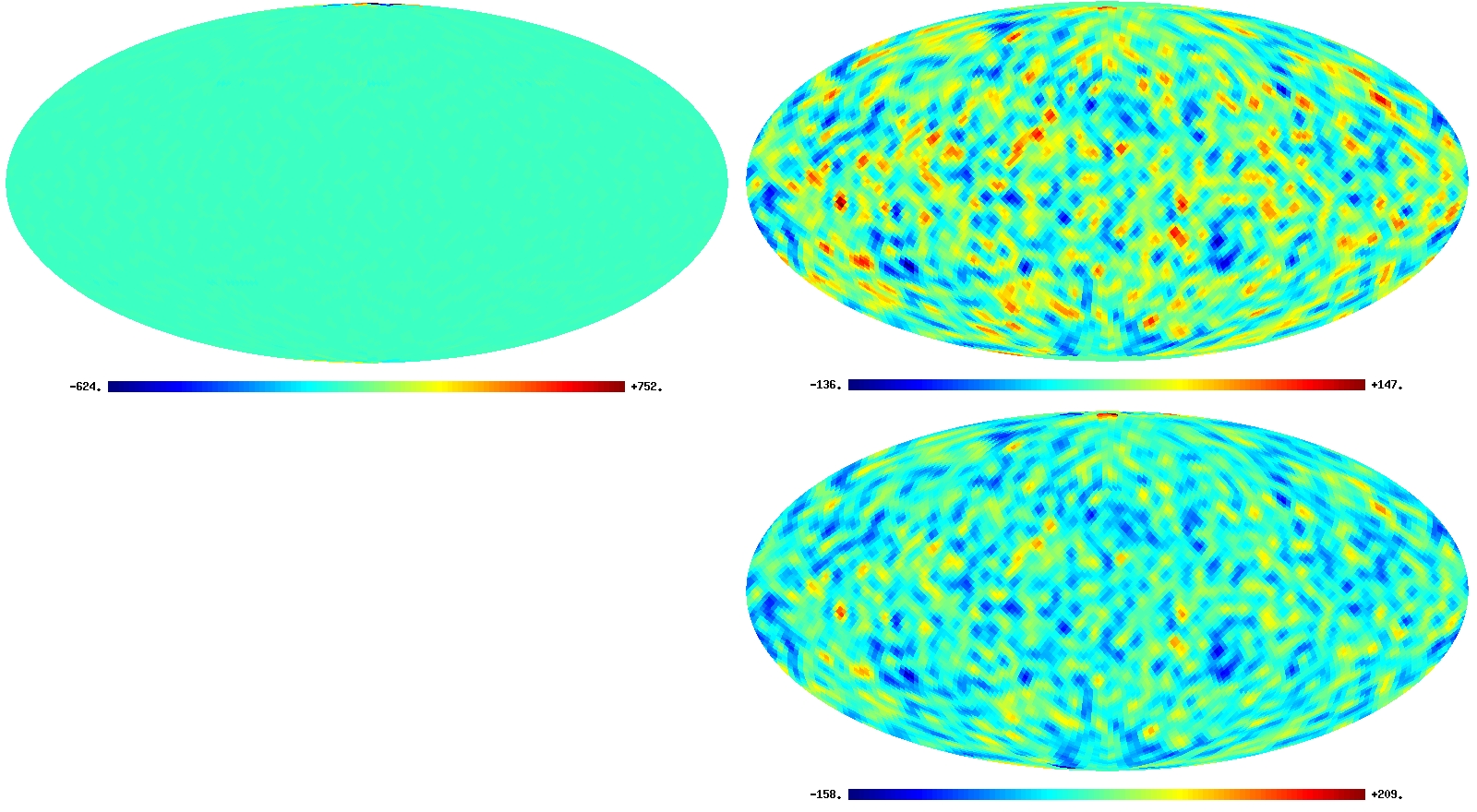}}\caption[]{Scalar $e$ maps calculated by MasQU at $n=2$, with various methods of dealing with the polar rings. Clockwise from top-left: no changes made, polar rings removed, 
rotation made in $\theta$. To remove any remaining large errors due to the closeness of a given ring to a pole, one can just increase the number of pixels treated with rotations.}\end{figure}

\noindent Rotation of a scalar function in $SO(3)$ is described by the Euler angles, most easily utilised by the rotation operator
\begin{equation}\mathcal{R}(a,b,c)=R_{z}(a)R_{y}(b)R_{z}(c)\end{equation}
where $a$, $b$, and $c$ are rotation angles. The equivalent rotation in the harmonic representation is written using the Wigner D-matrices:
\begin{equation}\mathcal{R}\phantom{}_{s}Y_{lm}=\sum_{m'=-l}^{l}D^{l}_{m'm}(\mathcal{R})\phantom{}_{s}Y_{lm'}.\end{equation}
The effect of rotation on $e$ and $b$ follows from their status as (pseudo-)scalar fields:
\begin{equation}\mathcal{R}e_{lm}Y_{lm}=e_{lm}\sum_{m'}D^{l}_{m'm}Y_{lm'}=\sqrt{\frac{(l-2)!}{(l+2)!}}a^{E}_{lm}\sum_{m'}D^{l}_{m'm}Y_{lm'}\end{equation}
such that the rotated and unrotated $E_{lm}$ harmonic coefficients are related by
\begin{equation}a^{E'}_{lm}=\alpha_{lm}(\mathcal{R})a^{E}_{lm}\end{equation}
and similarly for $a^{B}_{lm}$, where the $\alpha$ coefficient is straightforward to compute numerically. 

\section{The ``Pole Problem'' and Accurate Reconstruction of the Harmonic Coefficients}

\noindent We know that the construction of our bi-Laplacians is sensitive at the pole to error in the constituent derivatives, and that operating on rotated $Q$ and $U$ maps, either incrementally in $\phi$ or by $\pi/2$ in $\theta$, can solve this problem. Particularly for incremental $\phi$ rotations, it is important to accurately calculate the harmonic coefficients \begin{math}a_{lm}\end{math} of the original map in order to produce an accurate rotated map.
\\The HEALPix method of reconstructing the harmonic coefficients of a scalar field is an iterative procedure [22], manifested in the $map2alm$ subroutine: one starts with a zeroth-order estimator
\begin{equation}a_{lm}^{T,(0)}=\frac{4\pi}{N_{\mathrm{pix}}}\sum_{i}^{N_{\mathrm{pix}}}T(\Omega_{i})Y^{*}_{lm}(\Omega_{i})\end{equation}
and resums the coefficients to form a map $T^{(0)}$. The next step is to take the difference map \begin{math}\delta T^{(0)}=T-T^{(0)}\end{math} and compute the zeroth order harmonic coefficients of $\delta T^{(0)}$ in the same manner as the zeroth order of $T$, then iterate and sum over coefficients to form the $n^{\mathrm{th}}$-order approximation:
\begin{equation}a_{lm}^{T,(n)}\approx\frac{4\pi}{N_{\mathrm{pix}}}\left[\sum_{i}^{N_{\mathrm{pix}}}T(\Omega_{i})Y^{*}_{lm}(\Omega_{i})+\sum_{j=0}^{n-1}\sum_{i}^{N_{\mathrm{pix}}}\delta T^{(j)}(\Omega_{i})Y^{*}_{lm}(\Omega_{i})\right].\end{equation}
The optimal \begin{math}a_{lm}\end{math} sampling scale for map reconstruction is $l_{\mathrm{max}}=2N_{\mathrm{side}}$, with an optimal number of iterations being 3 according to the HEALPix software recommendations. While this is a quick and reasonable approximation, a numerical analysis finds that the convergence limit for the iterations is not suitable for the rotated sampling method in $\phi$.
\\In order to achieve accurate harmonic coefficients, we wish to perform the integral
\begin{equation}a_{lm}=\int F(\Omega)Y_{lm}^{*}(\Omega)d\Omega\end{equation}
For a HEALPix grid, the points in $\theta$ are described by [6]
\begin{equation}\begin{split}\cos\theta=\frac{4}{3}-\frac{2k}{3N_{\mathrm{side}}}\indent\mbox{ North equatorial belt, }N_{\mathrm{side}}\leq k\leq2N_{\mathrm{side}}
\\ \cos\theta=1-\frac{k^{2}}{3N_{\mathrm{side}}^{2}}\hspace{6.5mm}\mbox{ North polar cap, }1\leq k\leq N_{\mathrm{side}}\hspace{0.70in}
\end{split}\end{equation}
(and correspondingly in the southern hemisphere). Although this is globally irregular, the HEALPix \begin{math}\theta\end{math} points may be split into regular (equatorial) and irregular (polar cap) parts.
\\For the regular grid part one might choose a Newton-Cotes [17] method. Since the equatorial region is equally spaced, we can use the composite trapezoid rule for the poles and sum it with a single \begin{math}n_{\mathrm{equator}}\end{math}-point Newton-Cotes(NC) approximation at the equator.
\\If an integral is approximated by
\begin{equation}\int fdx=\sum_{i}w_{i}f_{i}dx,\end{equation}
where the nodes are equally spaced, then it can be shown that the NC weights can be found by solving 
\begin{equation}\left(\begin{array}{ccc}x_{1}^{0} &\cdots & x_{n}^{0} \\ \vdots & \ddots & \vdots \\x_{1}^{n} &\cdots & x_{n}^{n} \end{array}\right)\left(\begin{array}{c}w_{1}\\ \vdots\\ w_{n}\end{array}\right)=\left(\begin{array}{c}\frac{x_{n}-x_{1}}{2}\\ \vdots\\ \frac{x_{n}^{n+1}-x_{1}^{n+1}}{2^{n+1}} \end{array}\right).\end{equation}
Table II shows the results for this method on a regular grid, and Table III a comparison with Gaussian methods.
\begin{table}[here] \caption[]{Numerical test of the NC scheme; from Mathematica we have $\int_{-1}^{1}f(x)dx\approx1.71125$; $\Delta_{x}=b-a$.}
\begin{tabular}{|c||c||c||c|}
\hline No. of nodes & x & NC scheme & result 
\\ \hline 2 &-1,1 & \begin{math}\frac{\Delta_{x}}{2}(f_{0}+f_{1})\end{math}& 1.21306
\\ \hline 3 &-1,0,1 & \begin{math}\frac{\Delta_{x}}{2}(f_{0}+4f_{1}+f_{2})\end{math} &2.60653 
\\ \hline 4 &-1,-1/3,1/3,1 & \begin{math}\frac{3\Delta_{x}}{8}(f_{0}+3f_{1}+3f_{2}+4f_{3})\end{math} &2.1771 
\\ \hline 5 &-1,-1/2,1/2,1 & \begin{math}\frac{2\Delta_{x}}{45}(7f_{0}+32f_{1}+12f_{2}+32f_{3}+7f_{4})\end{math} &1.71047 
\\ \hline 6 &-1,-3/5,-1/5,1/5,3/5,1 & \begin{math}\frac{5\Delta_{x}}{288}(19f_{0}+75f_{1}+50f_{2}+50f_{3}+75f_{4}+19f_{5})\end{math} & 1.71082
\\ \hline 7 &-1,-4/6,-2/6,0,2/6,4/6,1 & \begin{math}\frac{\Delta_{x}}{140}(41f_{0}+216f_{1}+27f_{2}+272f_{3}+27f_{4}+216f_{5}+41f_{6})\end{math}&1.71128 
\\ \hline 
\end{tabular}\end{table}

\subsection{Gaussian quadrature on the HEALPix sphere}

\noindent Since we know how the regular and irregular HEALPix $\theta$ grid parts are constructed, one can then attempt a mixed NC and Gaussian scheme for the regular and irregular parts respectively, or even a Gaussian scheme for the whole sphere. Gaussian quadrature (in particular Gauss-Legendre quadrature) is used in GLESP to yield fast and highly accurate harmonic coefficients without recourse to iterative methods. The HEALPix points do not follow the abscissas of the Gauss-Legendre scheme, so we need to derive the general Gaussian scheme first. Starting from the integral [23]
\begin{equation}I=\int^{b}_{a}F(x)W(x)dx\approx\sum_{j=1}^{m}w_{j}f(x_{j})\end{equation}
and using an interpolating polynomial $L_{j}$ as given in Eq. (17), one finds
\begin{equation}w_{j}=\int^{b}_{a}\frac{L_{j}(x)W(x)}{x-x_{j}}dx\end{equation}
such that the relevant integral to solve is
\begin{equation}I=\int\frac{\prod_{i=1}^{m}(x-x_{j})}{x-x_{j}}dx.\end{equation}
A general solution to the above integral with $W(x)=1$ over $m$ points is
\begin{equation}I_{m}=\ln(x-x_{j})\prod_{i=1}^{m}(x_{j}-x_{i})+\sum_{z=1}^{m-1}x_{j}^{z}\left[\sum_{\alpha=0}^{z}(-1)^{\alpha}\frac{x^{z-\alpha}}{\alpha}S_{\alpha}\right],\indent S_{\alpha}=\sum^{\mathrm{combinations}}_{y_{0}\neq\cdots\neq y_{\alpha}}x_{y_{0}}\cdots x_{y_{\alpha}},\end{equation}
giving for example a 3-point solution
\begin{equation}I_{3}=\frac{x^{3}}{3}+\frac{x^{2}}{2}\left(x_{j}-\sum_{i=1}^{3}x_{i}\right)+x\left(x_{1}x_{2}+x_{1}x_{3}+x_{2}x_{3}+x_{j}^{2}-x_{j}\sum_{i=1}^{3}x_{i}\right)\end{equation}
where the log term disappears since we can only realistically sample at the pixel positions $i$.
\\The performance of this method is better than the NC method, and almost competitive with Gauss-Legendre (see Table II) --- although it can be shown that our general Gaussian scheme reduces to Gauss-Legendre at the requisite nodes.
\begin{table}[here] \caption[]{4-pt results summary for various methods}
\begin{tabular}{|c||c|}
\hline Method & value 
\\ \hline Exact & 1.71125 
\\ \hline Gauss-Legendre & 1.71122
\\ \hline Newton-Cotes (regular) & 2.1771
\\ \hline General Gauss Method (regular) & 1.7222
\\ \hline General Gauss Method (irregular) & 1.67456
\\ \hline
\end{tabular}\end{table}

\noindent There does however remain an important problem with this scheme: the number of combinations to find for, say, an $N_{\mathrm{side}}=32$ HEALPix map for a mixed Gauss and NC (Gaussian at the polar caps) and full Gaussian scheme respectively are
$\sum_{\alpha=1}^{30}31![(31-\alpha)!\alpha!]^{-1}$ and $\sum_{\alpha=1}^{126}127![(127-\alpha)!\alpha!]^{-1}$
which have prohibitively large time complexities.

\section{General Convolution Operators}

\noindent When calculating the residual for operation of $n$-point discrete derivatives in different frames (such as the Stokes tensor frame), or even for different spin-weight fields (such as if one has a vector field) it is useful to look at the more general convolution operator
\begin{equation}\phantom{}_{ss'}\mathcal{W}_{ll'mm'}=\int\partial_{\hat{n}}\left[\phantom{}_{s'}Y_{l'm'}(\Omega)\right]\phantom{}_{s}Y_{lm}(\Omega)d\Omega.\end{equation}
We briefly expound on some of the techniques available for such a calculation, without carrying out an in-depth example. One can first calculate the convolution operator for the derivatives in the analytic case by utilising the relation between the spin-weighted harmonics and the Wigner D-functions
\begin{equation}D^{l}_{-m,s}(\phi,\theta,\psi)=e^{im\phi}d^{l}_{-m,s}(\theta)e^{-is\psi}=(-1)^{m}\sqrt{\frac{4\pi}{2l+1}}\phantom{}_{s}Y_{lm}(\theta,\phi)e^{is\psi},\end{equation}
where the ``little-$d$'' function is
\begin{equation}\begin{split}d^{l}_{mm'}(\theta)=\sum_{r=m'-m}^{l+m}\frac{(-1)^{r}\sqrt{(l+m')!(l-m')!(l+m)!(l-m)!}}{(l+m-r)!r!(m'-m+r)!(l-m'-r)!}\left(\cos\frac{\theta}{2}\right)^{2l+m-m'-2r}\left(\sin\frac{\theta}{2}\right)^{m'-m+2r},\end{split}\end{equation}
and the angular momentum operators in $\mathbb{R}^{3}$ [16] are
\begin{equation}\begin{split}\hat{L}^{2}=-\left[\partial_{\theta}^{2}+\cot\theta\partial_{\theta}+\csc^{2}\theta(\partial_{\phi}^{2}-2\cos\theta\partial_{\psi\theta}+\partial_{\psi})\right]\hspace{1.23cm}
\\ \hat{L}_{z}=-i\partial_{\phi}\indent\hat{L}_{z'}=-i\partial_{\psi}
\indent\hat{L}_{\pm}=-\frac{1}{\sqrt{2}}e^{\pm i\phi}\left(\partial_{\theta}\pm i\cot\theta\partial_{\phi}\right)\end{split}\end{equation}
where it should be noticed that on $\mathbb{S}^{2}$, the operator $\hat{L}^{2}$ is equivalent to $D^{+}_{0}$. These operators have the power contributions
\begin{equation}\begin{split}\hat{L}^{2}D^{l}_{mm'}(\phi,\theta,\psi)=l(l+1)D^{l}_{mm'}(\phi,\theta,\psi)\hspace{1.42in}
\\ \hat{L}_{z}D^{l}_{mm'}(\phi,\theta,\psi)=-mD^{l}_{mm'}(\phi,\theta,\psi)\hspace{1.655in}
\\ \hat{L}_{z'}D^{l}_{mm'}(\phi,\theta,\psi)=-m'D^{l}_{mm'}(\phi,\theta,\psi)\hspace{1.61in}
\\ \hat{L}_{\pm}D^{l}_{mm'}(\phi,\theta,\psi)=\pm\sqrt{\frac{l(l+1)-m(m\mp1)}{2}}D^{l}_{m\mp1,m'}(\phi,\theta,\psi)\end{split}\end{equation}
which allows us to construct the analytic derivatives of the spin-harmonics using the recurrence relations of the Wigner terms (p.90 of [16]). As an illustrative example, the recurrence relation
\begin{equation}\begin{split}\sin\theta e^{\pm i\phi}D^{l}_{m\pm1,m'}(\phi,\theta,\psi)=\mp\frac{\sqrt{(l\pm m)(l\pm m+1)(l^{2}-m'^{2})}}{l(2l+1)}D^{l-1}_{mm'}(\phi,\theta,\psi)\\+\frac{m'\sqrt{(l\pm m)(l\pm m+1)}}{l(l+1)}D^{l}_{mm'}(\phi,\theta,\psi)\\ \pm\frac{\sqrt{(l\pm m)(l\pm m+1)[(l+1)^{2}-m'^{2}]}}{(l+1)(2l+1)}D^{l+1}_{mm'}(\phi,\theta,\psi)
\end{split}\end{equation}
shows that the $\partial_{\theta}$ operator redistributes power across multipoles $l$. To finish the evaluation of the analytic convolution operators, it is necessary to evaluate the integral
\begin{equation}I=\int\phantom{}_{s'}Y_{l'm'}(\Omega)\phantom{}_{s}Y_{lm}(\Omega)d\Omega.\end{equation}
This can be achieved by utilising the orthogonality relations of the little-$d$ functions
\begin{equation}\begin{split}\int^{\pi}_{0}d^{l}_{mm'}(\theta)d^{l'}_{mm'}(\theta)\sin\theta d\theta=\frac{2}{2l+1}\delta_{ll'}\end{split}\end{equation}
and those of the Jacobi polynomials $P_{l}^{(\alpha,\beta)}$ ([39], p.806), since
\begin{equation}d^{l}_{mm'}(\theta)=\sqrt{\frac{(l+m)!(l-m)!}{(l+m')!(l-m')!}}\left(\sin\frac{\theta}{2}\right)^{m-m'}\left(\cos\frac{\theta}{2}\right)^{m+m'}P_{l-m}^{(m-m',m+m')}(\cos\theta).\end{equation}
For discrete derivatives on a regular grid with sampling points separated by a length $\Delta$, the generalized convolution operator for derivatives in $\phi$ is trivial to compute. For the analysis of derivatives in $\theta$, we can use a perturbation expansion in the little-$d$ functions, $d^{l}_{mm'}(\cos(\theta+\Delta))$; by isolating the lowest order expansions in the sinusoidal terms in Eq. (B9)
\begin{equation}\begin{split}\left(\cos\left[\frac{\theta+\Delta}{2}\right]\right)^{2l+m-m'-2r}\left(\sin\left[\frac{\theta+\Delta}{2}\right]\right)^{m'-m+2r}\approx \left(\cos\frac{\theta}{2}\right)^{2l+m-m'-2r}\left(\sin\frac{\theta}{2}\right)^{m'-m+2r}
\\ \times(1+(m'-l-m+2r+l\cos\theta)\Delta\csc\theta)+\mathcal{O}(\Delta^{2})\end{split}\end{equation}
we see that the discrete derivatives in $\theta$ mix power across multipoles $s$, $l$ and $m$.

\end{document}